\documentclass[prb,twocolumn,superscriptaddress,showpacs,longbibliography]{revtex4-1}
\usepackage{graphicx}
\usepackage{epsfig}
\usepackage{amssymb,amsmath,amsfonts,hyperref}
\usepackage{wasysym}
\usepackage{latexsym}
\usepackage{eucal}
\usepackage{verbatim}
\usepackage[caption=false]{subfig}
\usepackage[section]{placeins}
\usepackage[normalem]{ulem}
\usepackage{color}

\newcommand{\be}{\begin{eqnarray}}
  \newcommand{\ee}{\end{eqnarray}}

\nocite{*}
\begin{document}
\title{Semiclassical theory for liquid-like behaviour of the frustrated magnet  $\mathrm{Ca}_{10}\mathrm{Cr}_{7}\mathrm{O}_{28}$}
\author{Sounak Biswas}
\affiliation{\small{Tata Institute of Fundamental Research, 1 Homi Bhabha Road, Mumbai 400005, India}}
\author{Kedar Damle}
\affiliation{\small{Tata Institute of Fundamental Research, 1 Homi Bhabha Road, Mumbai 400005, India}}
\begin{abstract}
  We identify the low energy effective Hamiltonian that is expected to describe the low
  temperature properties of the frustrated magnet $\mathrm{Ca}_{10}\mathrm{Cr}_{7}\mathrm{O}_{28}$. Motivated by the fact that this effective Hamiltonian has
  $S=3/2$ effective moments as its degrees of freedom, we use semiclassical
  spinwave theory to study the $T=0$ physics of this effective model and argue that singular spinwave fluctuations
  destabilize the spiral order favoured by the exchange couplings of this effective Hamiltonian. We also use
  a combination of classical Monte-Carlo simulations and molecular dynamics, as well as analytical approximations, to study
  the physics at low, nonzero temperatures. The results of these nonzero temperature calculations capture the liquid-like structure
  factors observed in the temperature range accessed by recent experiments. Additionally, 
  at still lower temperatures, they predict that a transition to nematic order in the bond energies reflects itself in the spin channel in the form of a crossover to a regime with large but finite correlation length for spiral spin correlations and a corresponding slowing
  down of spin dynamics.
\end{abstract}

\pacs{75.10.Jm}
\vskip2pc

\maketitle
\section{Introduction}
\label{intro}
At a phenomenological level, spin liquids are magnetic materials which avoid ordering down to the 
lowest temperatures studied, well below the temperature scale set by the exchange
interactions. This sets them apart from most other magnetic materials
which order at the temperature scale of the exchange interactions. This negative characterization of a spin liquid, although rooted in experimental phenomenology, is of limited utility
from a theoretical point of view.  A lot of theoretical effort over the years
has therefore been devoted to a more positive characterization of spin liquid
phases, in terms of topological order, emergent gauge structure, and fractional
excitations.\cite{Balents,SavaryBalents} 

Systems with geometrically frustrated antiferromagnetic interactions, which
result in a macroscopic degeneracy of low-energy configurations that
minimize the (classical) energy, are natural candidates
for spin liquid behaviour. One example is the frustrated magnet SCGO ($\mathrm{SrCr}_{9p}{\mathrm{Ga}}_{12-9p}{\mathrm{O}}_{19}$), which serves as a paradigmatic example
of a classical spin liquid, in which the observed behaviour can be explained in terms of
the macroscopic degeneracy of ground states of $S=3/2$ moments
on the SCGO lattice in the classical limit, with the effects of thermal fluctuations and
non-magnetic impurities also accounted for within this classical approximation.\cite{Limot,Schiffer_Daruka,Sen_Damle,Sen_Damle_PRB} Other
examples include minerals such
as Herbertsmithite and Volborthite, and  organic solids like
$\kappa$-(ET)$_2$Cu$_2$(CN)$_3$ , which are well-studied candidates for quantum spin liquid behaviour.\cite{Lee,Balents}

Recently, Balz {\em et. al.}\cite{Balz,Balz2} added to this list of candidates with a report of 
spin liquid behaviour in the compound $\mathrm{Ca}_{10}\mathrm{Cr}_{7}\mathrm{O}_{28}$. In $\mathrm{Ca}_{10}\mathrm{Cr}_{7}\mathrm{O}_{28}$, 
the spin $S=1/2$ $\rm{Cr}^{5+}$ ions form magnetically isolated Kagome bilayers.  
Using high field data on the one-magnon (single spin flip) excitation spectrum above the
fully-polarized ground state,  Balz {\em. et. al.}\cite{Balz,Balz2} have argued that the
magnetic Hamiltonian consists of nearest-neighbour Heisenberg exchange
couplings in each Kagome layer of the bilayer, as well as ferromagnetic exchange couplings between the two layers that make up each bilayer. A key feature of the exchange couplings extracted from their analysis is the following: In each bilayer, the up (down) pointing triangles of the lower (upper) Kagome layer host relatively large {\em ferromagnetic}
exchange couplings, while the down (up) pointing triangles of the lower (upper)
Kagome layer host significantly smaller antiferromagnetic exchange couplings roughly equal in magnitude to the ferromagnetic exchange interactions
that couple the upper and lower Kagome layers to each other (see Fig.~\ref{dbkagome}). One of the reasons for the recent
interest in $\mathrm{Ca}_{10}\mathrm{Cr}_{7}\mathrm{O}_{28}$ is the fact that
spin liquid behaviour is observed in spite of the dominant ferromagnetic couplings.\cite{Balz,Balz2}

Here, we provide an alternative theoretical perspective that relates the low temperature physics of $\mathrm{Ca}_{10}\mathrm{Cr}_{7}\mathrm{O}_{28}$
to the semiclassical large-spin limit of honeycomb lattice antiferromagnets with frustrating next-nearest
neighbour couplings. Our starting point is the following simple observation:
Since the dominant intralayer ferromagnetic couplings are at least three times larger in magnitude compared to the intralayer antiferromagnetic
and interlayer ferromagnetic couplings,\cite{Balz,Balz2} it should be possible to obtain a fairly accurate description of the low energy part of the spectrum by working with effective
$S=3/2$ degrees of freedom that represent the total spin of ferromagnetically
coupled up (down) pointing triangles of the lower (upper) Kagome layer in each bilayer (see Fig.~\ref{dbkagome}).
We expect this crucial simplification
to be valid below a temperature scale set by the magnitude of these dominant
intralayer ferromagnetic couplings. Since $S=3/2$ magnets can usually be described
in classical terms fairly well (except possibly at ultra-low temperatures which the experiments of Balz {\em et. al.} do not access), 
this observation immediately opens to
door to a semiclassical treatment\cite{Moessner_Chalker_prl,Moessner_Chalker_prb} of the problem. 

As will be clear below, the  pattern of exchange
couplings extracted by Balz {\em et. al.} from their analysis of the high-field
data implies that these $S=3/2$ degrees
of freedom can be thought of as occupying sites of a {\em honeycomb lattice} with nearest-neighbour ferromagnetic exchange couplings ($J_1$) and next-nearest-neighbour antiferromagnetic exchange couplings ($J_2$) of roughly equal magnitude. In our work here, we
perform a semiclassical analysis of the properties of this honeycomb lattice model, with a view towards understanding the liquid-like behaviour observed in experiments at not-too-low temperature.\cite{Balz,Balz2}
Our basic conclusion is that such a semiclassical description reproduces the observed liquid-like structure factors seen in the temperature range
accessed by recent experiments on $\mathrm{Ca}_{10}\mathrm{Cr}_{7}\mathrm{O}_{28}$. Additionally,
our results predict a lower temperature crossover to a regime with large but finite
correlation length for spiral spin correlations and a corresponding increase in spin autocorrelation times. This crossover occurs at roughly the same temperature at which
the bond energies are known to develop nematic order.\cite{Arun} This onset of nematicity in the bond energies is also related to the observed three-fold symmetry breaking phase transition seen in the work of Okumura {\em et. al.}~\cite{Okumura} in the classical model in
a different parameter regime of $J_2/J_1$.

The physical
picture that emerges from our analysis is as follows: The effective spin $S=3/2$ moments
can minimize their classical exchange energy by forming {\em spiral} states
at any wavevector ${\vec{q}}$ that falls on a one-dimensional locus ${\bf Q}_s$ in reciprocal space.  The leading $1/S$ corrections about any such classical spiral state labeled
by ${\vec{q}}$ consist of two bands of harmonic spinwave fluctuations. Including the
zero point energy of these spinwaves selects a spirals with a specific set of zone-boundary wavevectors that minimizes this leading $1/S$ correction to the ground state energy. However, the energy $E_{-}(\vec{k})$ of the lower band of spinwaves vanishes whenever $\vec{k}$ approaches any point on the entire one-dimensional locus of spiral wavevectors ${\bf Q}_s$ (in addition to vanishing at wavevector $\vec{k}=0$). Within this harmonic theory
of spinwave fluctuations, this vanishing of $E_{-}(\vec{k})$ on
the entire locus ${\bf Q}_s$ is crucially implicated in the {\em logarithmic divergence}
of the mean-square amplitude of transverse fluctuations about any such classical spiral state. 

This divergence of transverse fluctuations, reminiscent of the mechanism by which
long range antiferromagnetic order is destroyed by spin wave fluctuations at $T=0$ in {\em one-dimensional systems}, suggests (by analogy to this well-understood one dimensional case) that spiral order is likely destabilized by
spinwave fluctuations at $T=0$, although further analysis would be
needed to account for possible subtleties arising from anharmonic (higher order
in $1/S$) corrections to this picture. We return to a brief discussion of this
point towards the end of this article.

Of greater relevance to
the experiments of Balz {\em et. al.} is the effect of thermal fluctuations on
this incipient spiral order. Our results show that thermal fluctuations
lead, below a crossover temperature scale,  to a regime with a large but finite correlation length for spiral correlations of the spins at a particular set of
entropically-selected zone-boundary wavevector on the spiral locus ${\bf Q}_s$. Additionally, we
find a characteristic increase in the spin relaxation times below this crossover temperature. These crossovers in the spin channel take place at roughly the same temperature as the sharp onset of nematic
correlations in the bond energies studied in the work of Mulder {\em et. al.}\cite{Arun}. In this low temperature regime, a large but
finite correlation length for spiral spin correlations thus coexists with nematicity in the bond energies. This relatively simple 
theoretical picture complements the more sophisticated pseudo-fermion
functional renormalization group analysis employed by Balz {\em et. al.} in their
own theoretical analysis of the underlying microscopic model of $S=1/2$ spins on the Kagome bilayer.
Most of the inelastic
neutron scattering results of Balz {\em et. al.} are at temperatures above this crossover.
In this regime, our calculations yield a liquid-like structure factor similar
to these experimental results. 

The rest of this paper is organized as follows. In Sec.~\ref{effec_model} we introduce the microscopic model Hamiltonian extracted from high-field data on this Calcium Chromate compound,\cite{Balz,Balz2} and identify the effective Hamiltonian
that governs the behaviour of the effective spin $S=3/2$ degrees of freedom that represent
the low energy degrees of freedom.
In Sec.~\ref{largen}, we carry out a large-$N$ study of this effective model within the
classical approximation ({\em i.e.} treating the $S=3/2$ spins as fixed-length vectors
of magnitude $S$), and calculate correlation
functions and structure factors to leading order in large-$N$. In Sec.~\ref{qspinw}, motivated by our large-$N$ results, we construct a degenerate set of spiral ground states
({\em Luttinger-Tisza spirals}) for the classical system, with spiral ordering wavevectors
$\vec{q}$ lying on a one-dimensional locus ${\bf Q}_s$ in reciprocal space, and study the effect of quantum-mechanical spin-wave fluctuations about these ground states to
leading order in the $1/S$ expansion. In Sec.~\ref{class_sel}, we study the effect of thermal fluctuations on the degenerate manifold of ground states in the classical limit. 
In Section~\ref{numerics}, we carry out a combined Monte Carlo-Molecular Dynamics study of the statics and dynamics of the effective model of
classical spins identified in Sec.~\ref{effec_model}, and present numerical results for
the temperature dependence of structure factor, specific heat, susceptibility and relaxation time.  We close with a brief discussion of some outstanding issues in Sec.~\ref{discussion}.

\section{The effective model} 
\label{effec_model}
\begin{figure}[t]
  \centering
  \includegraphics[width=8.6 cm]{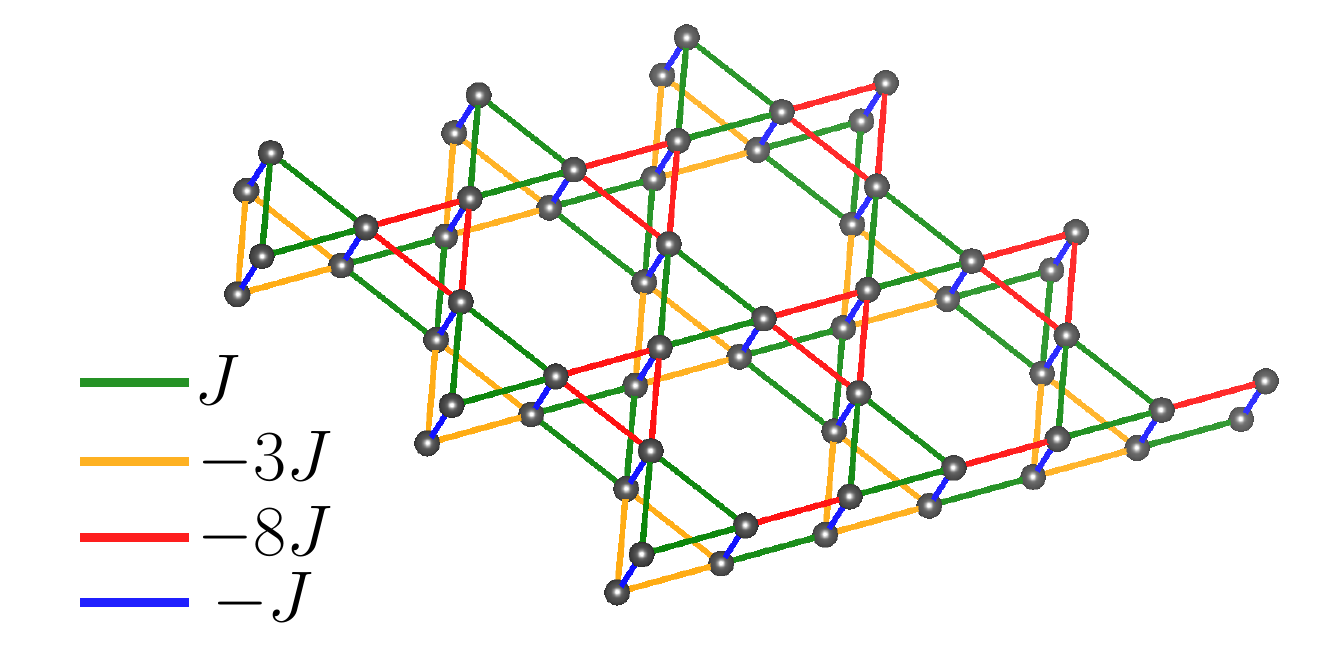}
  \caption{(Color online) The $\mathrm{Cr}^{5+}$ ions in $\mathrm{Ca}_{10}\mathrm{Cr}_{7}\mathrm{O}_{28}$ form a Kagome bilayer structure, as reported in Ref.~\onlinecite{Balz}. Each layer has antiferromagnetic couplings $\simeq J=0.1 \rm{meV} $ as well as much stronger ferromagnetic couplings , as indicated in the figure, while the interlayer couplings are again antiferromagnetic with magnitude $\simeq J=0.1 \rm{meV} $. Green and red links constitute
the upper Kagome layer, while green and yellow links constitute the lower Kagome layer. Blue links denote interlayer couplings. This figure
  has been created using VESTA.~\cite{Vesta}}
  \label{dbkagome}
\end{figure}

The crystal structure and magnetic properties of the magnetic insulator $\mathrm{Ca}_{10}\mathrm{Cr}_{7}\mathrm{O}_{28}$ were studied recently by Balz. et al.\cite{Balz,Balz2} using
x-ray diffraction and inelastic neutron scattering methods as well as thermodynamic
measurements. The magnetic $\rm{Cr}^{5+}$ ions ($S=1/2$) were found to form Kagome bilayers,
with each bilayer magnetically isolated from the next by the absence of exchange pathways. Using inelastic neutron scattering at high magnetic fields, it was possible to map out the excitation spectrum
of single spin-flip ``magnon'' excitations above the fully-polarized high-field ground state.
The form of the microscopic Hamiltonian governing the dynamics of the $S=1/2$
Kagome bilayers was deduced from fits to this data in conjunction with thermodynamic measurements. This analysis yielded the best-fit Hamiltonian
\begin{equation}
  H({\vec{S_{i}}})=\sum_{ij}M^{\rm bare}_{ij}\vec{S}_{i} . \vec{S}_{j} .
  \label{Heisen}
\end{equation}
The isotropic Heisenberg exchange couplings that make up the matrix $M^{\rm bare}_{ij}$ above may be described as follows:
Up-pointing (down-pointing) triangles of the lower (upper) Kagome layer in
each bilayer consist of three spins strongly coupled to each other by strong
ferromagnetic bonds of magnitude $J^{F}_{ll}$ ($J^{F}_{uu}$), while the exchange couplings that constitute the
links of down-pointing (up-pointing) triangles in the lower (upper) Kagome
layer are antiferromagnetic, with a significantly smaller magnitude $J^{AF}_{ll}$ ($J^{AF}_{uu}$). Additionally, spins directly
above one another are connected by a ferromagnetic exchange interaction that
couples the two layers of each Kagome bilayer. This has magnitude $J^{F}_{ul}$.
To within the error
bars quoted by Balz {\em et. al.}, $J^{AF}_{ll} \simeq J^{AF}_{uu} \simeq J^{F}_{ul} \equiv J$,
$J^{F}_{ll} \simeq 3J$, $J^{F}_{uu} \simeq 8J$, with $J \simeq 0.1 \rm{meV}$. The magnetic lattice, as well as this pattern of exchange couplings, is displayed in Fig.~\ref{dbkagome}.

\begin{figure}[t]
  \centering
  \includegraphics[width=8.6 cm]{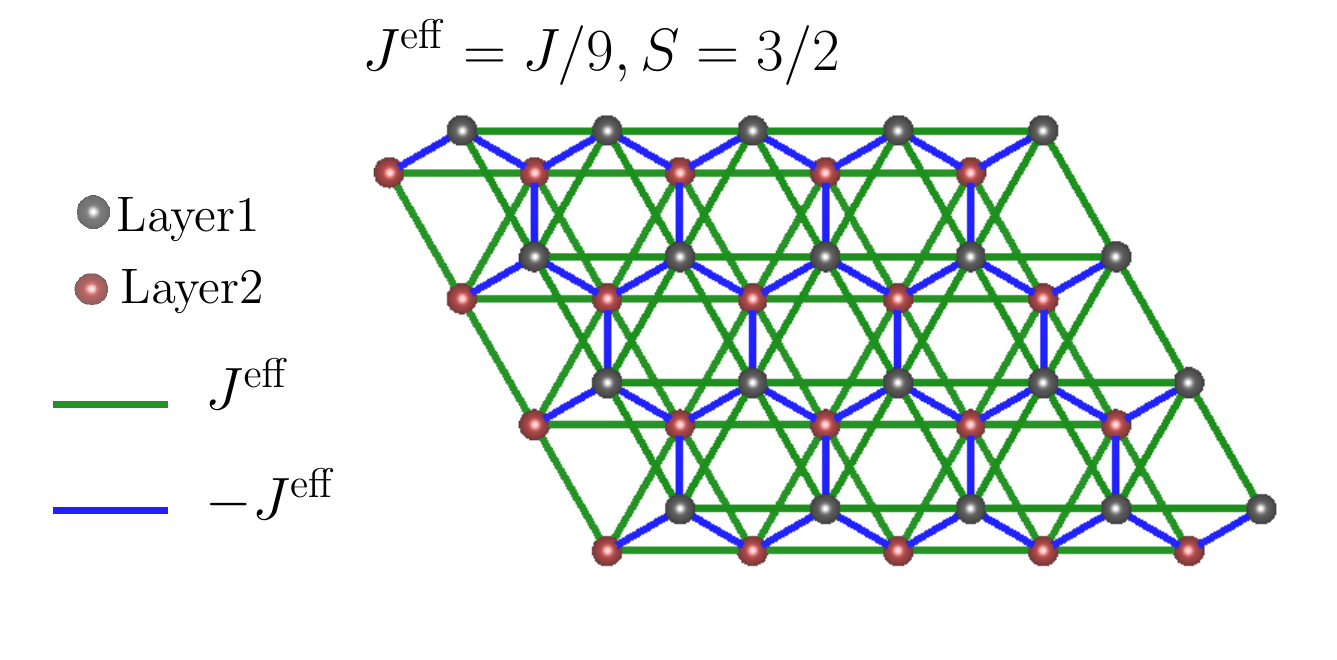}
  \caption{The low energy effective Hamiltonian has spin $S=3/2$ moments on a bilayer triangular lattice, with antiferromagnetic intra-layer couplings and ferromagnetic inter-layer couplings as shown. The magnitude of all couplings in this effective Hamiltonian is $J^{\mathrm{eff}}=J/9$, where $J$ is the microscopic in-plane antiferromagnetic coupling between the Cr$^{5+}$ spins. This is equivalent to a honeycomb lattice with nearest-neighbour ferromagnetic couplings
  and next-nearest-neighbour antiferromagnetic couplings. When written in terms of unit vector $\hat{n}$ instead of vectors of length $S=3/2$,
  the effective model has couplings of magnitude $J^{\mathrm{eff}} S^2 \simeq 290 \rm{mK} \simeq 0.025 \rm{meV}$. Energies (frequencies) and temperatures are measured in units of this energy scale in all subsequent figures. This figure
  has been created using VESTA.~\cite{Vesta}}

  \label{effec}
\end{figure}

As already noted by Ref.~\onlinecite{Balz}, the ferromagnetic exchange couplings $J^{F}_{ll}$ and $J^{F}_{uu}$ dominate over the antiferromagnetic couplings $J^{AF}_{ll}$, $J^{AF}_{uu}$, and $J^{F}_{ul}$, being at least three times larger than these
antiferromagnetic couplings. Our starting point is the observation that low energy eigenstates are expected
to be built from states in which the three spins coupled by $J^{F}_{ll}$ ($J^{F}_{uu}$)
in the lower (upper) Kagome layer are in a total spin $S_{\rm tot} =3/2$ state.
This strongly suggests that the low energy physics should be described
by an effective Hamiltonian written in terms
of spin $S=3/2$ moments that represent such strongly ferromagnetically coupled
triangles. These strongly ferromagnetically coupled triangles in each Kagome
layer thus form a triangular lattice of $S=3/2$ moments, which represent states in the
total spin $=3/2$ multiplet of the three spin $S=1/2$ moments coupled together by the strong ferromagnetic couplings acting within each such triangle. To obtain the effective interaction of these $S=3/2$ effective moments with each other to leading order in the ratios of subleading couplings to the dominant ferromagnetic couplings, we must project these subleading couplings into the subspace of states obtained by restricting to the total spin $S=3/2$ multiplet of each strongly coupled triangle. Performing this projection, we see that the $S=3/2$
effective moments are coupled to each other by nearest neighbour antiferromagnetic Heisenberg exchange interactions of magnitude $J^{\mathrm{eff}}=J/9$. In addition, to the same
accuracy, the effect of the interlayer coupling $J^{F}_{lu}$ is to introduce an effective
ferromagnetic interlayer coupling of the same magnitude $J^{\mathrm{eff}}=J/9$, which couples
the two triangular layers of $S=3/2$ moments. This is
shown in Fig.~\ref{effec}. For the rest of this paper,
we work with this effective model, which is expected to capture the physics correctly below a temperature
scale set by the strong ferromagnetic couplings in each layer.

We note that this  effective model of $S=3/2$ moments on a bilayer triangular lattice is equivalent to a  $J_{1}-J_{2}$ Heisenberg model on a honeycomb lattice, with nearest neighbour ferromagnetic interactions $J_1=J^{\mathrm{eff}}=J/9$ (corresponding to the interlayer coupling between the two triangular layers that make up a bilayer) and next-nearest-neighbour
antiferromagnetic interactions $J_2$ of the same magnitude (corresponding to the
antiferromagnetic interaction between spin $S=3/2$ moments on the same
triangular layer). The spin-$S$ $J_1-J_2$ Heisenberg model on the
honeycomb lattice has been the subject of several previous studies
in the context of materials in the 
BaM$_{2}$(XO$_{4}$)$_{2}$ (M=Co, Ni; X= Pt, As) family and the Bi$_3$M$_4$O$_12$(NO$_3$) family(M= Mn,V,Cr).~\cite{Rastelli, Fouet,Arun,Okumura,Arun2,Mattsson,PFFRG}, and we will
make contact with these studies when we discuss our results.  In our classical molecular dynamics and Monte Carlo studies, we choose to represent
the classical $S=3/2$ moments by unit vectors; this necessitates a rescaling
of the exchange couplings by a factor of $|S|^2$, so that we work with
a model of unit vectors interacting with an exchange coupling of strength $J^{\rm{eff}}|S|^2=J/4$. For convenience, we quote all numerical values  in units of $J/4$ or $(J/4)^{-1}$ in the rest of this paper ($J/4$ corresponds to approximately $0.025 \rm{meV}$ or $290 {\rm mK}$).

\section{Large-$N$ study}
\label{largen}
The problem of finding classical groundstates given a pattern of exchange couplings is a constrained minimization problem. Instead of
attacking it right away, we use the
large-$N$ approximation,\cite{Ma} whereby we generalize from the $O(3)$ degrees of freedom (in terms of which
we write the classical limit of our spin Hamiltonian) to $O(N)$ vectors obeying the constraint 
$ \vec{\phi}_{i}^{2}=N$ on each site $i$, and then use the $N\rightarrow \infty$ solution to approximate the behaviour at $N=3$. This follows the path laid out by similar calculations for other frustrated classical spin systems.\cite{ Sen_Damle,Sen_Damle_PRB,Wang, Garanin_Canals1, Garanin_Canals2, Isakov} 

When working within the large-N approximation, we choose to represent the $S=3/2$ moments $\vec{S}$ as classical ($c$-number) vectors $\vec{\phi}$ of length $\sqrt{3}$ (instead of unit vectors that are a more convenient representation for our combined Monte Carlo and molecular
dynamics computations). Thus we write
$\vec{S} = \sqrt{3} \vec{\phi}/2$. In this language, the
Hamiltonian is written as
\begin{equation}
  H(\{\hat{n}_{i} \})=(1/3)\sum_{ij}\vec{\phi}_{i}\cdot \vec{\phi}_{j}M_{ij}.
  \label{unitvecH}
\end{equation}
Here, $M_{ij}$ is the pattern of couplings depicted in Fig.~\ref{effec} with the exchange couplings $J^{\rm{eff}}$ rescaled by a factor of $|S|^{2}=9/4$, so that elements of $M_{ij}$ have magnitude $J^{\rm{eff}}S^2$ as alluded to in the end of the previous section. The additional prefactor of $1/3$ 
in Eq.~\eqref{unitvecH} of course accounts for the rewriting in terms of vectors $\vec{\phi}$ of
length $\sqrt{3}$.The lattice of Fig.~\ref{effec} is a triangular Bravais lattice with a two-site unit cell representing the two layers of the original system. As noted in the previous section, it is equivalent, as far as the connectivity (not geometry) is concerned,  to a honeycomb lattice with nearest and next-nearest-neighbour couplings. In Eq.~\eqref{unitvecH} and all subsequent discussion we adopt the convention
that $i,j$ are composite indices comprising of the Bravais lattice site with coordinate $\vec{r}_i$, and a sublattice (layer) index $\alpha$ ($\alpha=1,2$). When inessential, we suppress the sublattice indices in what follows.

The expression 
for the partition function in the large-$N$ limit becomes
\begin{equation}
  Z\propto \int \prod_{i}d\vec{\phi}_{i}\exp(-\beta H)\prod_{i}\delta
  \Big(\vec{\phi}_{i}^{2}-N\Big).
  \label{pathint}
\end{equation}
Using $\delta(x)=\int \exp(i \lambda x)$, and the expression for the Hamiltonian in Eq.~\eqref{unitvecH},
we can write the partition function (Eq.~\eqref{pathint}) as

\begin{align}
  \nonumber Z\propto \int & \mathcal{D}[\lambda]\mathcal{D}[\vec{\phi}] \exp(iN \sum_{i}\lambda_{i}) \\
  &\times \exp \Big(-\sum_{ij}\vec{\phi}_{i}\cdot \vec{\phi}_{j}(\frac{\beta}{3} M_{ij}+i\lambda_{i}\delta_{ij})\Big),
  \label{pathint2}
\end{align}
where we have used $\mathcal{D}[\lambda]=\prod_{i}d\lambda_{i}$, and $\mathcal{D}[\vec{\phi}]=\prod_{i}d\vec{\phi}_{i}$. The $\lambda_{i}$ integrals can be performed exactly using the
fact that the saddle-point approximation becomes exact in the $N\rightarrow \infty$ limit. 
Setting all $\lambda_{i}=\lambda$, as is appropriate for a saddle-point that respects
all lattice symmetries, one has
\begin{equation}
  Z\propto \int\mathcal{D}[\vec{\phi}]\exp(-\frac{\beta}{3} \sum_{ij}\vec{\phi}_{i}\cdot \vec{\phi}_{j}(M_{ij}+\overline{\lambda}\delta_{ij}))
  \label{sadHam}
\end{equation}
where $\overline{\lambda}$ is the saddle point value of $3i\lambda/\beta$, self-consistently determined 
by the equations
\begin{equation}
  \langle \phi_{i}^{2}\rangle_{\overline{\lambda}}=1
  \label{sadeqn}
\end{equation}
for each site $i$. Here $\phi_i$ is a scalar field that represents any one component of $\vec{\phi}_i$.
\begin{figure}[t]
  \includegraphics[width=4.6 cm]{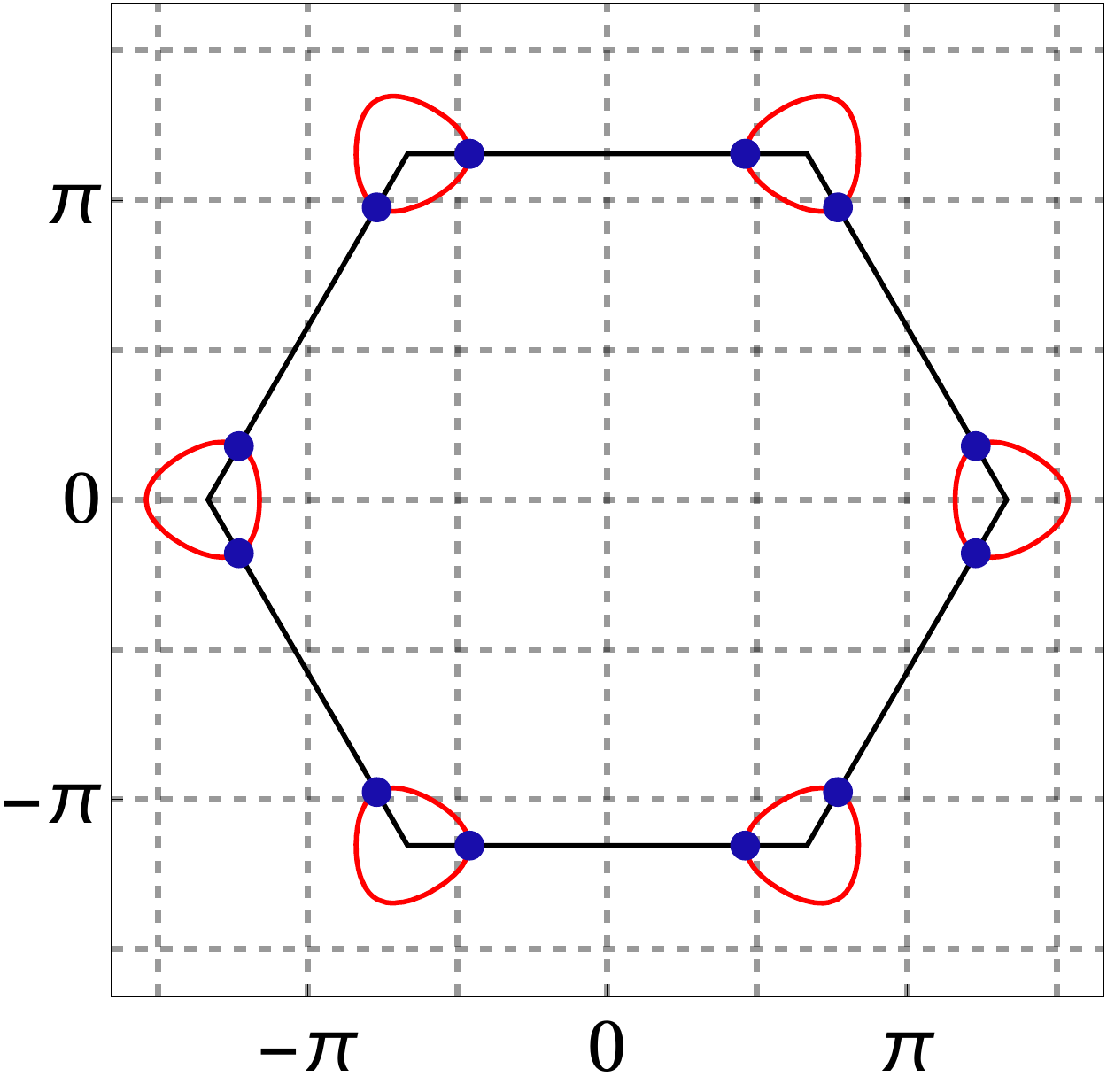}
  \caption{Wave vectors labeling degenerate ground states lie on the locus $\mathbf{Q_s}$ marked in red. The hexagon marks the first Brillouin zone of the triangular Bravais lattice, and the $x$ and the $y$ axes refer to components of $\vec{q}$ in the $\hat{x}$ and $\hat{y}$ directions. Also marked are the points at the zone boundaries of the first Brillouin zone, which are selected by quantum fluctuations as well as thermal fluctuations (see main text for details).}
  \label{Minima}
\end{figure}

To find the lowest energy configurations that dominate the large-$N$ path integral in the low temperature limit, we diagonalize the 
saddle point Hamiltonian matrix $M_{ij}+\overline{\lambda}\delta_{ij}$. We do this in Fourier space, where it is block diagonal. Our lattice is a triangular Bravais lattice with a two site unit cell. We introduce a sublattice index in the subscript of the scalar fields to write  
$\phi_{\alpha}(\vec{k})= \sum_{r_i}\phi_{\alpha,r_i}\exp(-i\vec{k} \cdot \vec{r}_i)$. Here and in all subsequent discussion, wavevectors
are measured in units of $a^{-1}$ and positions in units of $a$, where $a$ is the lattice spacing of the underlying triangular Bravais lattice, 
which we estimate to be
$\simeq 5.35 \rm{A}$ from the more precise measurements of the crystal structure given in Ref.~\onlinecite{Balz,Balz2} (small distortions from perfect Kagome bilayer geometry have been ignored in arriving at our estimate).

Here, $\alpha$ denotes the sublattice and the sum runs over unit cells. 
Expressing vectors in terms of their components along the principal axes $\hat{e}_1$ and $\hat{e}_2$ of the triangular lattice (with
$\hat{e}_1 \cdot \hat{e}_2 = -1/2$), we have
\begin{align}
  \sum_{i,j}n_{i} M_{ij}n_{j}=\frac{1}{L^2}\sum_{\vec{k}}\Phi(\vec{k})^{\dagger}M(\vec{k})\Phi(\vec{k}) ,\\
  \Phi^{\dagger}(\vec{k})=(\phi^{*}_{1}(\vec{k}),\phi^{*}_{2}(\vec{k})),\\
  M(\vec{k})=\frac{1}{2}J^{\rm{eff}}|S|^{2}	\left(
  \begin{array}{cc}
    \Delta  & -K^{*} \\
    -K & \Delta  \\
  \end{array}
  \right) \; ,
\end{align}
where $\Delta= 2(\cos(k_1)+\cos(k_2)+\cos(k_1+k_2))$ and $K=(1+\exp(i k_1)+\exp(i k_1+ i k_2))$.
The eigenvalues are given by $E^{\pm}(\vec{k})=\frac{1}{2}J^{\rm{eff}}|S|^{2}(\Delta \pm \sqrt{\Delta+3})$. 

These eigenvalues describe two dispersive bands. The lower band $E^{-}(\vec{k})$ 
has degenerate band minima labeled by wave vectors $\vec{q}$ such that
\begin{equation}
  2 ( \cos(q_1)+\cos(q_2)+\cos(q_1+q_2))= -11/4
  \label{minima}
\end{equation}
The solutions of this equation lie on a locus $\mathbf{Q_s}$ shown in Fig.~\ref{Minima}. It is worthwhile to compare this
degeneracy with what one has as a result of large-N calculations for other frustrated systems which are known to 
exhibit spin-liquid behaviour: $\mathrm{SrCr}_{9p}{\mathrm{Ga}}_{12-9p}{\mathrm{O}}_{19}$, 
in which the lattice is a pyrochlore slab with nearest neighbour interactions, has a seven
site unit cell and seven bands, of which the lowest three are flat. The pyrochlore lattice itself has, within this approximation, four bands, out of which the lower two are flat.\cite{Garanin_Canals2} Herbertsmithite, where the spins are on a Kagome lattice, has
three bands, out of which the lowest is flat.~\cite{Garanin_Canals1} Within large-N
such flat bands are usually signatures of liquid-like
behaviour. 
Our line-degeneracy is reminiscent of Volborthite~\cite{Wang} where the spins lie on a distorted Kagome 
lattice and the lower band minima form a one dimensional degenerate subspace. 
\begin{figure}[t]
  \subfloat[T=0.35]{\includegraphics[width= 0.48\columnwidth]{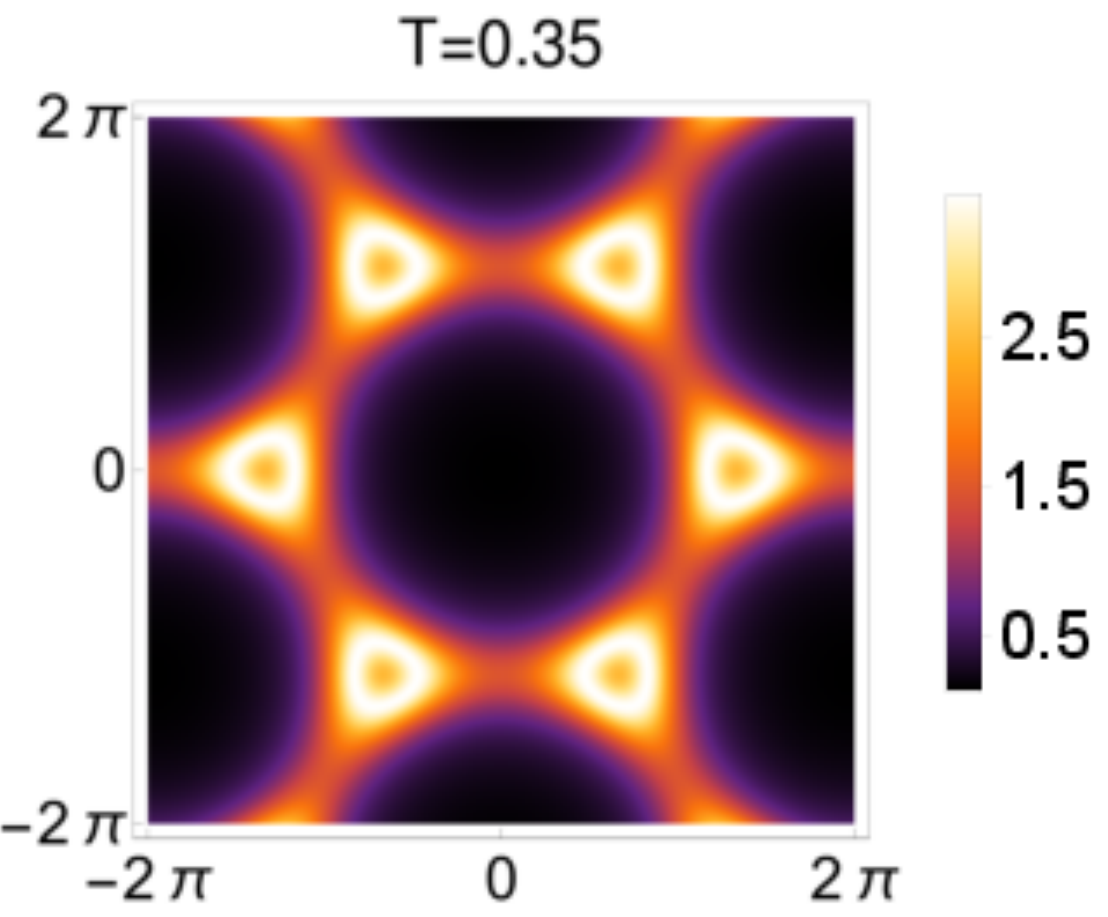}}
  \subfloat[T=3.50]{\includegraphics[width=0.48\columnwidth]{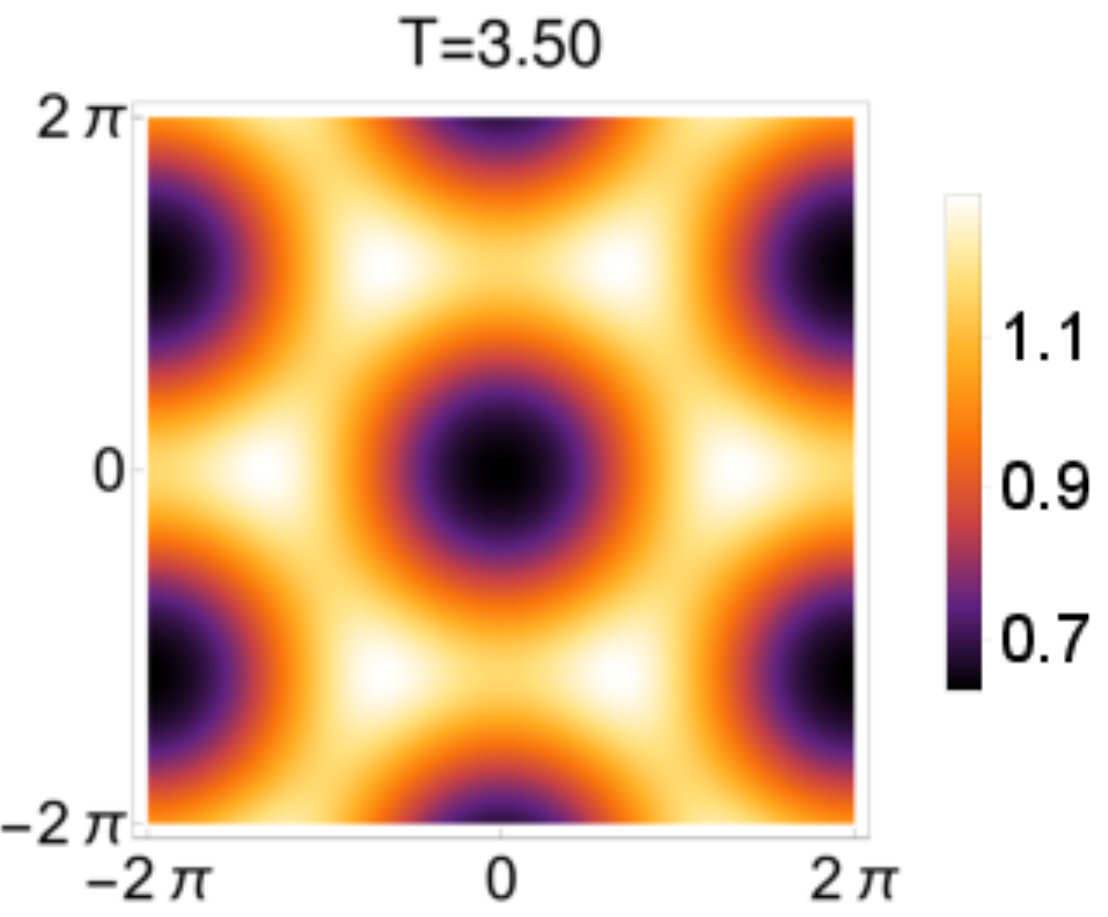}}
  \caption{In-plane momentum dependence (with out of plane momentum set to zero) of correlation functions of spins in the same plane, $G_{11}(\vec{k})=G_{22}(\vec{k})$ (Eq.~\eqref{gkk2}), computed within the large-N approximation at temperatures (a) $0.35 (J/4) \simeq 100 \rm{mK}$ and 
  (b) $3.50 (J/4) \simeq 1 \rm{K}$ ($(J/4) \simeq 290 \rm{mK}$). The lower temperature results show clear features
associated with the tendency towards spiral order.}
\label{lNcorr}
\end{figure}

The eigenvectors for any point $\vec{q}$ on this locus
are given by $\phi^{\pm}=(\frac{1}{\sqrt{2}})(1,\mp\exp(i \theta_{\vec{q}}))$, where $\theta_{\vec{q}}$ is determined by
\begin{align}
  \nonumber \cos(\theta_{\vec{q}})=2(1+\cos(q_1)+\cos(q_1+q_2)) \\
  \sin(\theta_{\vec{q}})=2(\sin(q_1)+\sin(q_1+q_2))
  \label{theta}
\end{align}
Note that the equation of the locus $\mathbf{Q_s}$ guarantees that this pair of equations for $\theta_{\vec{k}}$ has a legitimate solution.

Next, we calculate spin correlations in this large-$N$ approximation by numerically solving Eq.~\eqref{sadeqn} to obtain $\overline{\lambda}(\beta)$ and
using this value to determine the equal time correlation function in momentum space.
For a system of $L\times L$ unit cells, this is given by
\begin{align}
  \label{gkk1}
  \langle \phi_{\alpha}(\vec{k})\phi_{\beta}(-\vec{k'})\rangle=L^{2}\delta_{\vec{k},\vec{k'}}G_{\alpha,\beta}(\vec{k}),\\
  \label{gkk2}
  G_{11}(\vec{k})=G_{22}(\vec{k})= \frac{3}{\beta}\frac{ \Delta+\overline{\lambda}(\beta)}
  {(\Delta+\overline{\lambda}(\beta))^{2}-(\Delta +3)},\\
  \label{gkk3}
  G_{12}(\vec{k})=G_{21}^{*}(\vec{k})= \frac{3}{\beta}\frac{ K}
  {(\Delta+\overline{\lambda}(\beta))^{2}-(\Delta +3)}.
\end{align}
\begin{figure}[t]
  \centering
  \subfloat[T=0.35]{\includegraphics[width= 0.48\columnwidth]{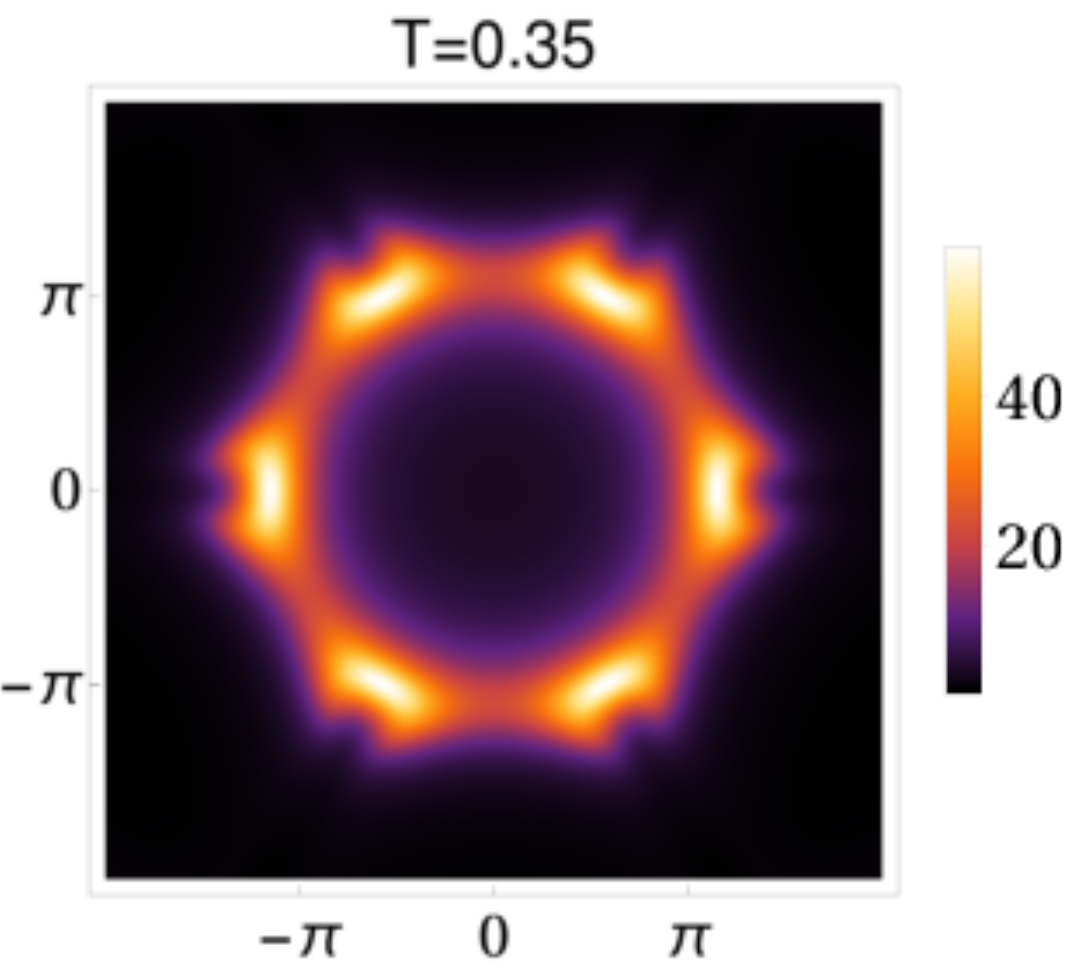}}
  \subfloat[T=3.50]{\includegraphics[width= 0.48\columnwidth]{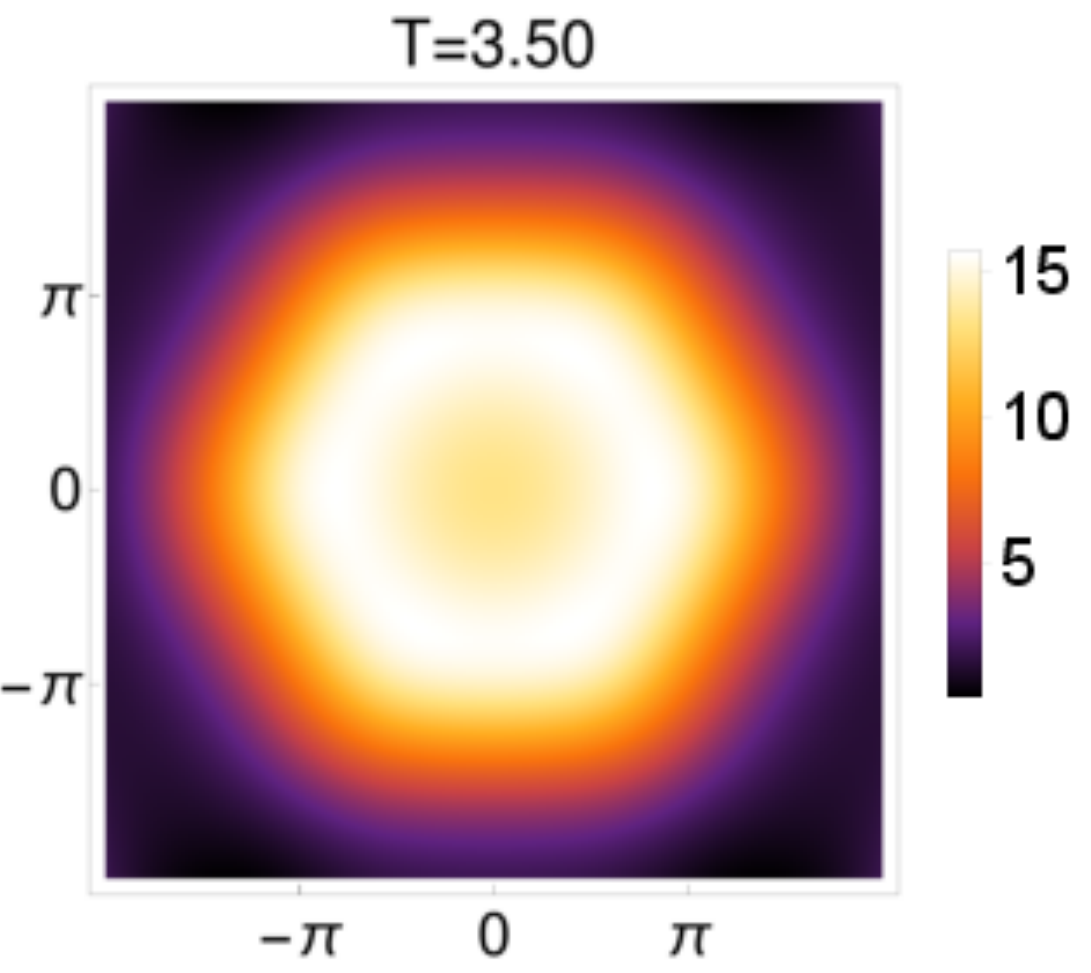}}
  \caption{In plane momentum dependence (with out of plane momentum set to zero) of the equal time structure factor of spins, $\mathcal{S}(\vec{k})$ (Eq.~\eqref{sfac}) within the large-N approximation
  at two values of temperature: (a) $T= 0.35 (J/4) \simeq 100 \rm{mK}$ and 
  (b) $T= 3.50 (J/4) \simeq 1 \rm{K}$ ($J/4 \approx 290 \rm{mK}$). We note that form factors partially smear out, but do not eliminate the ``spiral features'' seen earlier (Fig.~{\protect{\ref{lNcorr}}}) in the intra-plane spin correlations at the lower temperature.}
  \label{largeNsfac}
\end{figure}
In Fig.~\ref{lNcorr}, we show the momentum-space correlation functions of spins in the same plane, $G_{11}(\vec{k})$, for two temperatures $T=100 mK$
and $T=1 K$ relevant to the
experiments performed in Ref.~\onlinecite{Balz}. 
One can also calculate the equal time spin structure factor within this approximation by using these results to compute
\begin{equation}
  \mathcal{S}(\vec{k})=\frac{1}{L^2}\langle \lvert \phi_{1}( \vec{k})f_{1}(\vec{k}) + \phi_{2}(\vec{k})f_{2}(\vec{k}) \rvert^{2} \rangle
  \label{sfac}
\end{equation}
where the subscripts denote the sublattice as before, and $f_1(\vec{k})$ and $f_2(\vec{k})$ are the form-factors 
for the bound $S=3/2$ degrees of freedom corresponding to triangular plaquettes of ferromagnetically 
coupled spin $1/2$ moments (see Appendix~\ref{formfacs}). In Fig.~\ref{largeNsfac}, we show the large-$N$ results for the equal
time structure factors at the same temperatures. The lower temperature scans at $T=100 mK$ clearly show features associated
with the tendency towards spiral order, although there is clearly no true long range order possible in this two dimensional system.
We also note that the form factors partially smear out these ``spiral features'', making them harder to observe in the equal time
spin structure factor (as opposed to the intra-plane correlation function displayed earlier).

\section{Spin-wave theory at $T=0$}
\label{qspinw}
From the large-$N$ ground states obtained in the previous section, we may construct physical ground states of three-component vectors of magnitude $|S|=3/2$.
Since  the eigenvectors of the exchange-coupling matrix $M_{ij}$ have the same magnitude on
both sublattices, it is possible to use these eigenvectors to construct valid classical ground states for the $S=3/2$ spins. These are the `Luttinger-Tisza' spiral ground-states~\cite{Luttinger},
obtained by making appropriate linear combinations of the eigenvectors $\phi^{\pm}$:
\begin{align}
  \begin{split}
    \vec{S}^{\rm GS}_{i}=|S|\hat{n}^{\rm GS}_i, (\hat{n}^{\rm GS}_i)^2=1,\\
    \hat{n}^{\rm{GS}}_{1,r_i}= \Big(\cos(\vec{q}.\vec{r}_i)\hat{z}+\sin(\vec{q}.\vec{r}_i)\hat{x}\Big),\\
    \hat{n}^{\rm{GS}}_{2,r_i}=\Big( \cos(\vec{q}.\vec{r}_i-\theta_{\vec{q}})\hat{z}+\sin(\vec{q}.\vec{r}_i-\theta_{\vec{q}})\hat{x}\Big),
  \end{split}
  \label{LT}
\end{align}
where $\vec{q}$ belongs to the ground-state manifold obtained from large-N results in
Eq.~\eqref{minima} and we have used the explicit representation of the composite index $i$ in terms of $(\alpha,r_i)$, where $\alpha$ is the sublattice index and $r_i$ is the coordinate of the underlying triangular Bravais lattice. These classical ground states are related (by a spin flip on one sublattice) to those constructed by Mulder {\em et. al.} in their study of the $S=1/2$ honeycomb
lattice $J_1$-$J_2$ with both couplings antiferromagnetic.\cite{Arun}

Now, we look at whether quantum fluctuations lift the degeneracy of the manifold of spiral ground states in
\eqref{LT}, and whether they render such spiral ordering unstable.
Although higher order corrections in $1/S$ (anharmonic corrections to the leading harmonic spinwave theory) are outside the scope of our
analysis, the leading order results may be expected to already be
fairly reliable for spin $S=3/2$.  Some
of our results in Sec.~\ref{qspinw} were obtained earlier in a different context in Ref~\onlinecite{Rastelli},
and are reproduced here in the interests of a self-contained presentation.
Our calculations are also analogous to similar spin-wave calculation by Mulder et al.~\cite{Arun} for the case of \emph{antiferromagnetic} $J_1$, although there is no canonical transformation that connects the two problems,
and the leading order spin-wave corrections (and the semiclassical spin dynamics) are therefore not the same although the classical ground states are closely related.

We consider spin-wave fluctuations about a spiral ordered state of Eq.~\eqref{LT} labeled by the wave vector $\vec{q}$ belonging to the degenerate groundstate locus  $\mathbf{Q_s}$ given by Eq.~\eqref{minima}. First, we rotate the local $\hat{z}$ 
axis to point along the spins in the spiral ordered state given by ~\eqref{LT}. This rotation transforms a 
generic quadratic term of our Heisenberg Hamiltonian in the following way.
\begin{align}
  \begin{split}
    \vec{S}_{i}. \vec{S}_{j}\rightarrow  & S^{y}_{i}S^{y}_{j} +(S^{z}_{i}S^{z}_{j} + S^{x}_{i}S^{x}_{j}) \cos(\omega_{i,j}) \\
    &  +(S^{z}_{i}S^{x}_{j} - S^{x}_{i}S^{z}_{j}) \sin(\omega_{i,j}). 
  \end{split}
\end{align}
where $\omega_{i,j}$ is given by
\begin{align}
  \begin{split}
    \omega_{i,j}=\omega_{\alpha,r_i;\beta,r_j}= \vec{q}.(\vec{r}_i-\vec{r}_j) \text{ for } \alpha=\beta, \rm{ and}\\
    \omega_{1,r_i;2,r_j}=-\omega_{2,r_j;1,r_i}=\vec{q}.(\vec{r}_i-\vec{r}_j)-\theta_{\vec{q}} \\
  \end{split}
\end{align}
Here $\theta_{\vec{q}}$ is defined
in Eq.~\eqref{theta}, and we have explicitly expressed the composite indices $i$ and $j$ in terms of the sublattice index $\alpha,\beta=1,2$ and the unit cell position coordinates $r_i,r_j$.
Next, we choose the spin quantization axis along the local $\hat{z}$ axis defined above and make a transformation
to Holstein-Primakoff bosons $b_{\alpha,i},b_{\alpha,i}^{\dagger}$, in effect making the substitutions $S^{z}\rightarrow S-b^{\dagger}b$, $S^{+}\rightarrow \sqrt{2S}b$,
and $S^{-}\rightarrow \sqrt{2S}b^{\dagger}$ (correct to quadratic order). We then expand the resulting expansions to leading order in $1/S$, 
again keeping terms only up to quadratic order in the boson creation and annihilation operators, to obtain a non-interacting spin-wave Hamiltonian 
$H_{\mathrm{SW}}(\vec{q})$.

To  diagonalize the spin-wave Hamiltonian, we transform to Fourier
space as 
$b_{\alpha}(\vec{k})=
\sum_{i}b_{\alpha,i}\exp(i\vec{k}.\vec{r}_i)$ , $\alpha$ labeling the sublattice, to obtain
\begin{align}
  H_{SW}(\vec{q})=& E^{\rm{GS}} +\frac{|S|}{L^2} \sum_{\vec{k}}'\mathbf{b^{\dagger}}(\vec{k}) \mathbf{M} (\vec{q},\vec{k})\mathbf{b}(\vec{k}) -2 a(\vec{q},\vec{k}),\\
  \mathbf{b^{\dagger}}=&(b_{1}^{\dagger}(\vec{k}),b_{2}^{\dagger}(\vec{k}),b_{1}(-\vec{k}),b_{2}(-\vec{k})).
  \label{hpbosons}
\end{align}
Here, $\sum^{'}$ denotes a sum over half of the Brillouin zone. The expressions for $a(\vec{q},\vec{k})$ and for the matrix $\mathbf{M}(\vec{q},\vec{k})$ are given in Appendix~\ref{spinw}. $E^{\rm{GS}}$ is the spiral ground-state energy independent of $\vec{k}$ and $\vec{q}$,  given in terms of the connectivity matrix $M_{ij}$ by

\begin{equation}
  E^{\rm{GS}}=\sum_{i,j}\hat{S}^{\rm{GS}}_{i}(\vec{q})M_{ij}S^{\rm{GS}}_{j}(\vec{q}),
  \label{e_gs}
\end{equation}
for any $\vec{q}$ in the spiral groundstate manifold $\mathbf{Q_s}$ defined by Eq.~\eqref{minima}.
The quadratic spin-wave Hamiltonian $H_{\mathrm{SW}}(\vec{q})$ can be diagonalized by making a canonical 
transformation to  Bogoliubov quasiparticles $\gamma_{\pm}(\vec{k})$, which preserve  the bosonic commutation
relations $[\gamma_{\mu}(\vec{k}),\gamma^{\dagger}_{\nu}(\vec{k'})]=\delta_{\vec{k},\vec{k'}} \delta_{\mu,\nu}$ ($\mu, \nu = \pm$).\cite{Colpa} In terms of the Bogoliubov quasiparticles, one can write
\begin{align}
  \label{bogol}
    & H_{SW}(\vec{q})= E_{\mathrm{GS}}+E^{0}(\vec{q})\\ \nonumber
    +&\frac{|S|}{L^2}\sum_{\vec{k}\in BZ,\sigma=\pm}E^{\rm{SW}}_{\sigma}(\vec{q},\vec{k})\gamma_{\sigma}^{\dagger}(\vec{k})\gamma_\sigma(\vec{k}). 
\end{align}
The spin-wave dispersions $E^{\rm{SW}}_{\pm}(\vec{q},\vec{k})=E^{\rm{SW}}_{\pm}(\vec{q},\vec{-k})$ are detailed in Appendix~\ref{spinw}. We note that
the lower band $E^{\rm{SW}}_{-}(\vec{k})$ has  zero energy modes at the spiral wave-vectors lying on the locus $\mathbf{Q_s}$ defined in Eq.~\ref{minima}, apart from a 
Goldstone mode at $k=0$.
The $\vec{q}$-dependent zero point energy of spin-wave fluctuations $E^{0}(\vec{q})$ is given by
\begin{equation}
  E^{0}(\vec{q})=\frac{|S|}{L^2}\sum_{\vec{k}}^{'}\Big(E^{\rm{SW}}_{+}(\vec{q},\vec{k})+E^{\rm{SW}}_{-}(\vec{q},\vec{k})-2a(\vec{q},\vec{k})\Big) \; .
  \label{zeropoint}
\end{equation}
To obtain the state favoured by spin-wave fluctuations, we minimize the zero-point energy  $E^{0}(\vec{q})$ in Eq.~\eqref{zeropoint}
over the 
classical ground state spiral wave vectors given by Eq,~\eqref{minima}. We find that, $E^{0}(\vec{q})$
is minimized for 
\begin{equation}
  (q_1,q_2)=\Big(\arccos(1/8),\pi-\arccos(3/4)\Big)
  \label{qgs}
\end{equation}
and the other wave vectors related by lattice symmetries.
Therefore, within non-interacting spin-wave theory, quantum fluctuations favour
the spiral states given by Eq.~\eqref{qgs} and other wavevectors
related by lattice symmetries. The wave-vectors favoured by quantum fluctuations lying within the first Brillouin
zone are shown in Fig.~\ref{Minima}.
\begin{figure}[t]
  \centering
  \includegraphics[width=\columnwidth]{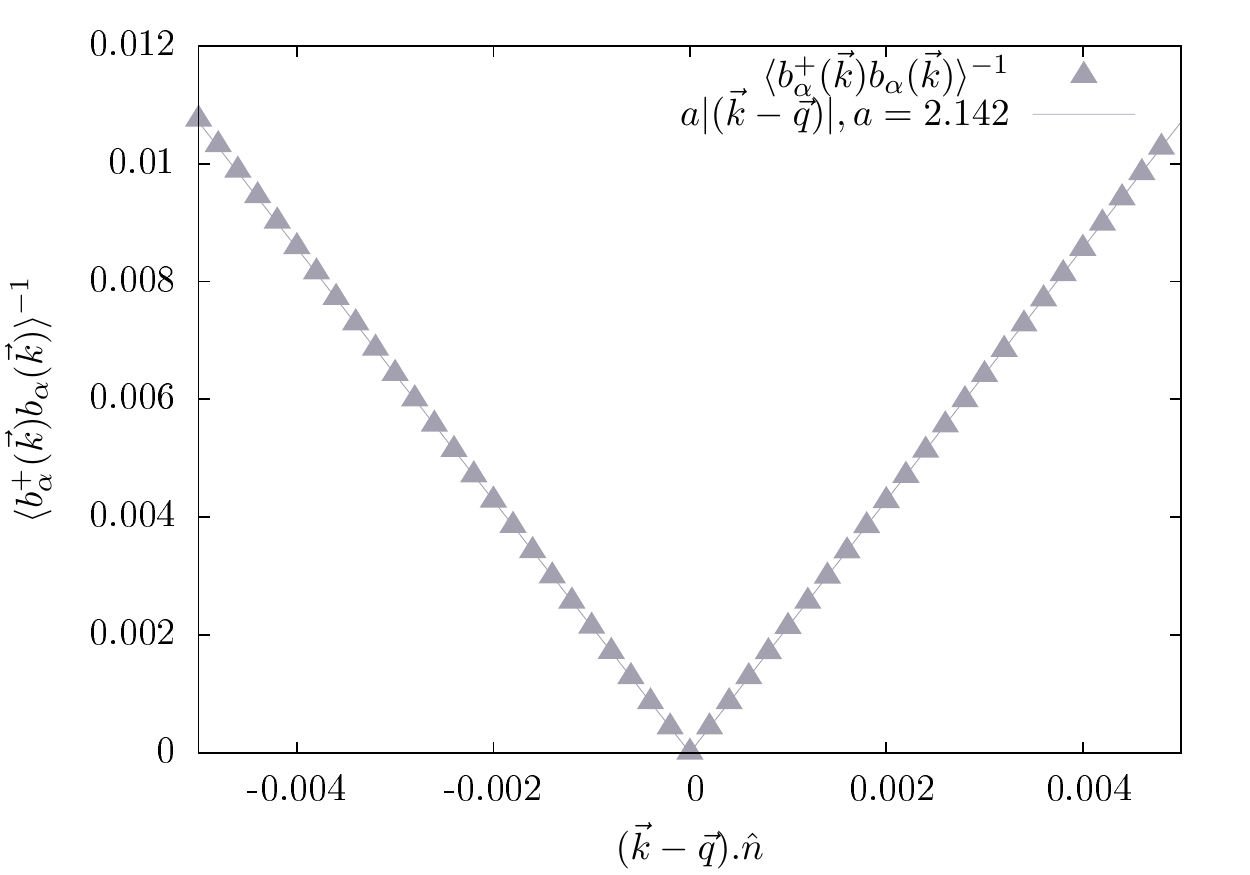}
  \caption{The expectation value $\langle b^{+}_{\alpha}(\vec{k})b_{\alpha}(\vec{k})\rangle^{-1}$ calculated within leading-order spin wave theory. The plotted wave vectors are along $\hat{n}$, the local normal at a generic point on the locus $\mathbf{Q_s}$ of the degenerate spiral wave-vectors given by Eq.~\eqref{minima}. The solid line is a fit to the form $a |\vec{k}-\vec{q}|$, with $a=2.142$. This linear behaviour, being generic along the spiral locus, signifies an instability of the spiral order to transverse fluctuations (see main text for details).}
  \label{linear}
\end{figure}

The Mermin-Wagner theorem rules out order at any finite temperature. The question of whether 
the system orders at zero temperature can be studied within spin-wave theory by looking 
at the expectation value of magnetization about the local $\hat{z}$ axis : 
\begin{equation}
  \frac{1}{2L^2} \langle \sum_{\alpha,i}S^{z}_{\alpha,r_i}\rangle = \frac{1}{2L^2} \sum_{\vec{k},\alpha}(S-\langle b^{+}_{\alpha}(\vec{k})b_{\alpha}(\vec{k})\rangle).
\end{equation}
A small expectation value of the Holstein-Primakoff boson number $(1/2L^2)\sum_{\vec{k},\alpha}\langle b^{+}_{\alpha}(\vec{k})b_{\alpha}(\vec{k})\rangle$
would imply that the spiral ground state is stable to transverse fluctuations. We numerically evaluate $\langle b^{+}_{\alpha}(\vec{k})b_{\alpha}(\vec{k})\rangle^{-1}$ 
and find that it
vanishes on the wave vectors belonging to the spiral manifold given by Eq,~\eqref{minima}. For small deviations perpendicular to the locus of degenerate spiral wave-vectors $\mathbf{Q_s}$,
we find that $\langle b^{+}_{\alpha}(\vec{k})b_{\alpha}(\vec{k})\rangle^{-1} \propto |\vec{k}-\vec{q}|$ where $\vec{q}$ is any location on the spiral manifold. For  a particular spiral wave vector $\vec{q}$, this linear dependence is shown in Fig.~\ref{linear}. We have checked that this linear behaviour does not depend on the location of wave vector $\vec{q}$ on the spiral manifold given by Eq.~\ref{minima}.  
This linear behaviour renders the integral $(1/2L^2)\sum_{\vec{k},\alpha}\langle b^{+}_{\alpha}(\vec{k})b_{\alpha}(\vec{k})\rangle$
logarithmically divergent in the thermodynamic limit. Within leading order spin-wave theory, we thus find that transverse fluctuations destabilize spiral order.
We note that the spiral order suffers the same fate in the system with antiferromagnetic inter-layer couplings~\cite{Arun}, even though the spin-wave dispersions
are different ( In this case, $\langle b^{+}_{\alpha}(\vec{k})b_{\alpha}(\vec{k})\rangle^{-1}$ is linear in perpendicular deviations $|\vec{k}-\vec{q}|$ with a different proportionality constant).
We note that the role of higher order terms in the
$1/S$ needs to be analyzed to obtain a more definite prediction regarding the fate of the system. 
In spite of this caveat regarding the ultimate fate of the system, this analysis does strongly suggest that spiral order, favoured by
the pattern of exchange couplings in the system, is destabilized
due by singular spinwave fluctuations, possibly opening the door to $T=0$ spin-liquid behaviour. Another competing possibility is bond-energy nematic order of the type
predicted for the $S=1/2$ case in the work of Mulder {\em et. al.}\cite{Arun}

\section{Classical fluctuations about spiral ground-states}
\label{class_sel}
Having studied the effect of quantum fluctuations on classical ground states in Sec.~\ref{qspinw}, 
we now look at the effect of thermal fluctuations. Our method follows the one used in Ref.~\onlinecite{Balents_diamond} in the analysis of the spinel MnSc$_2$S$_4$. A similar
calculation has been reported earlier for a different regime of $J_2/J_1$.\cite{Okumura}

In this section, we work with configurations of unit-vectors $\hat{n}$, such that $\vec{S}=|S|\hat{n}$.
We consider fluctuations about the configuration $\hat{n}^{GS}(\vec{q})$, where $\hat{n}^{GS}(\vec{q})$ is the unit-vector configuration describing the spiral groundstate $\vec{S}^{GS}(\vec{q})$ defined in Eq.~\eqref{LT} and $\vec{q}$ belongs to the degenerate groundstate locus $\mathbf{Q_s}$. 
The configuration $\hat{n}_{i}$ can be written in terms of fields $\vec{\epsilon_{i}}$ describing fluctuations from $\hat{n}^{GS}_{i}$ as
\begin{equation}
  \hat{n}_{i}=\vec{\epsilon_{i}}+\hat{n}_{i}^{GS}(\vec{q})\sqrt{1-\vec{\epsilon_{i}}^{2}}.
  \label{fluc1}
\end{equation}
The fluctuation fields $\vec{\epsilon}_i$ satisfy $\vec{\epsilon_{i}}.\hat{n}_{i}^{GS}=0$, and are always constrained to obey $\vec{\epsilon_{i}}\leq 1$. Together with the  form of Eq.~\eqref{fluc1}, these conditions explicitly preserve the unit vector constraint on the spins. In terms of the fluctuation fields $\vec{\epsilon}$, one can write the partition function as
\begin{align}
  Z= & \int\mathcal{D}[\hat{n}]\exp(-\beta H)\\
  = & \int\mathcal{D}[\epsilon]\exp(-\beta H)\prod_{i}(1-\vec{\epsilon_{i}}^{2})^{-\frac{1}{2}},
  \label{flucpart}
\end{align}
where we have put in the expression for the Jacobian of the transformation from the $\hat{n}_{i}$ to the $\vec{\epsilon_{i}}$ fields.
The fluctuation fields $\vec{\epsilon_{i}}$ can be further decomposed into scalar fields
$\pi_{i}$ and $\rho_{i}$ describing fluctuations in and out of the plane of
the spiral  as
\begin{equation}
  \vec{\epsilon_{i}}=\rho_{i}\hat{y}+\pi_{i}(\hat{y}\times\vec{S_{i}}^{GS}(\vec{q})).
  \label{fluc2}
\end{equation}
We absorb the Jacobian into the exponential and express the partition function of 
Eq.~\eqref{flucpart} in terms of the scalar fields
$\rho$ and $\pi$ using Eq.~\eqref{fluc1} and Eq.~\eqref{fluc2}. Expanding in these  fields and keeping 
terms up to quadratic order in $\rho$ and $\pi$ gives us the leading order partition function of
small fluctuations about an ordered spiral ground state:
\begin{align}
  \nonumber  Z= & \int \mathcal{D}[\pi]\mathcal{D}[\rho]\exp\Big(-\mathcal{S}(\pi,\rho)\Big)\\
  \mathcal{S} = &\beta \sum_{ij}( \rho_i J_{ij} \rho_{j}+ \pi_i K_{ij} \pi_j)-\frac{1}{2}\sum_i (\rho_i^2+\pi_i^2),
  \label{flucpart1}
\end{align}
where the matrices $J_{ij}$ and $K_{ij}$ are defined in terms of the connectivity matrix $M_{ij}$ of 
Eq.~\eqref{unitvecH} and the spiral ground-state energy $E^{\rm{GS}}$ (Eq.~\eqref{e_gs}) as
\begin{align}
  \nonumber J_{ij}= & M_{ij}-E^{GS}\delta_{ij},\\
  K_{ij}= & (M_{ij}-E^{GS}\delta_{ij})\vec{S_{i}}^{GS}(\vec{q}) \cdot \vec{S_{j}}^{GS}(\vec{q}).
  \label{newconnect}
\end{align}
We note that the in-plane fluctuation matrix $K_{ij}$ has two bands as expected. The lower band has
zeros exactly at the spiral wave vectors belonging to the degenerate groundstate locus $\mathbf{Q_s}$ given by Eq.~\eqref{minima}, i.e., it has a one-dimensional 
subspace of soft fluctuation modes (or zero modes), just like the connectivity matrix $M_{ij}$,
apart from a zero mode at $\vec{k}=0$.

Now, one can ask what states among the degenerate manifold of spiral ground states are 
entropically selected at nonzero, but low temperatures. In this regime, one can drop the 
temperature independent Jacobian terms in the partition function of small fluctuations 
about the ordered spiral state $\vec{S_i}^{\rm{GS}}$. The fluctuation fields $\rho$
and $\pi$ can be integrated out to give 
\begin{align}
  \nonumber Z(\vec{q})= & \int\mathcal{D}[\pi]\mathcal{D}[\rho]\exp\Big(-\beta\sum_{ij}( \rho_i J_{ij} \rho_{j}+ \pi_i K_{ij}(\vec{q}) \pi_j)\Big)\\
  \propto & \; \; \; \; \; \mathrm{det}(\beta J)^{-1/2} \mathrm{det}(\beta K(\vec{q}))^{-1/2}
  \label{flucpart2}
\end{align}
where we have explicitly shown the dependence on the spiral wave vector $\vec{q}$. To find the states selected entropically, we minimize the free energy $F(\vec{q})= -T\log(Z(\vec{q})))$ over the 
manifold of spiral states given by Eq.~\eqref{minima}. The free energy, up to additive constants 
independent of temperature or the spiral wave-vector $q$, is given by
\begin{equation}
  F(\vec{q})=\frac{1}{\beta}\mathrm{Tr}(\log(\beta J))+\frac{1}{\beta}\mathrm{Tr}(\log(\beta K(\vec{q}))).
  \label{free}
\end{equation}
The first term is independent of the spiral wave-vector $\vec{q}$, and cannot break the degeneracy of the groundsates given by Eq.~\eqref{minima}. As detailed in Appendix~\ref{app_class}, the trace in the second term can be easily calculated in the
Fourier basis, where $K(\vec{q})$ is block-diagonal. In this way, we find that the states selected by small fluctuations 
at small nonzero temperatures are the same as the ones selected at zero temperature by non-interacting 
spin-waves, i.e., ones at the edges of the first Brillouin zone, given by Eq.~\eqref{qgs} 
and shown in Fig.~\ref{Minima}.
  \section{Numerical Study}
  \label{numerics}
  In this section we undertake a combined Monte Carlo-Molecular Dynamics study of the 
  classical effective spin $3/2$ model described earlier.
  \subsection{Method}
  \label{method}
  To study equilibrium properties and equal time correlation functions, we use Monte Carlo simulations. While
  embedded cluster algorithms are available for continuous spin systems ~\cite{Wolf,Hasenbusch}, the extremely 
  frustrated nature of the low temperature configurations of this model render these inefficient.
  Therefore, following Refs.~\onlinecite{Lee_Young} and \onlinecite{Pixley_Young}, we use three single-spin updates :
  a) \emph{Over-relaxation} moves are energy-conserving micro-canonical sweeps, which reflect the 
  spin of each site about the effective magnetic field, b) \emph{Heat-bath moves} to equilibrate each spin in the external exchange field of
  its neighbours, and  c) \emph{Parallel Tempering} , which exchanges, with acceptance
probability that obeys detailed balance, entire configurations between two independent
  simulations run at slightly different temperatures. More details on these update schemes can be found in Ref.~\onlinecite{Lee_Young}. For completeness, we have also documented the details relevant to our implementation in Appendix~\ref{MCapp}.
\begin{figure}[t]
  \centering
  \subfloat[T=0.20, L=64]{\includegraphics[width= 0.48\columnwidth]{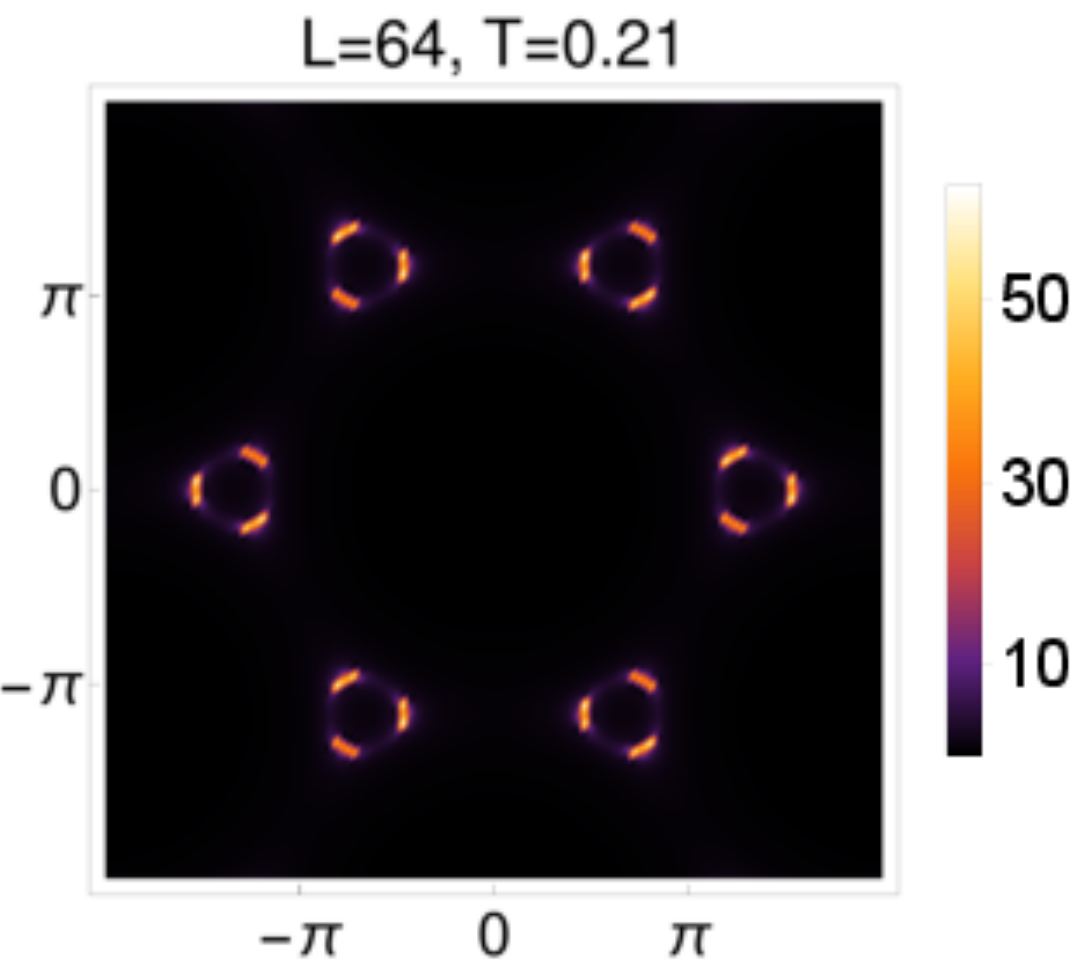}{\label{MCT0.20}}}
  \subfloat[T=0.22, L=64]{\includegraphics[width= 0.48\columnwidth]{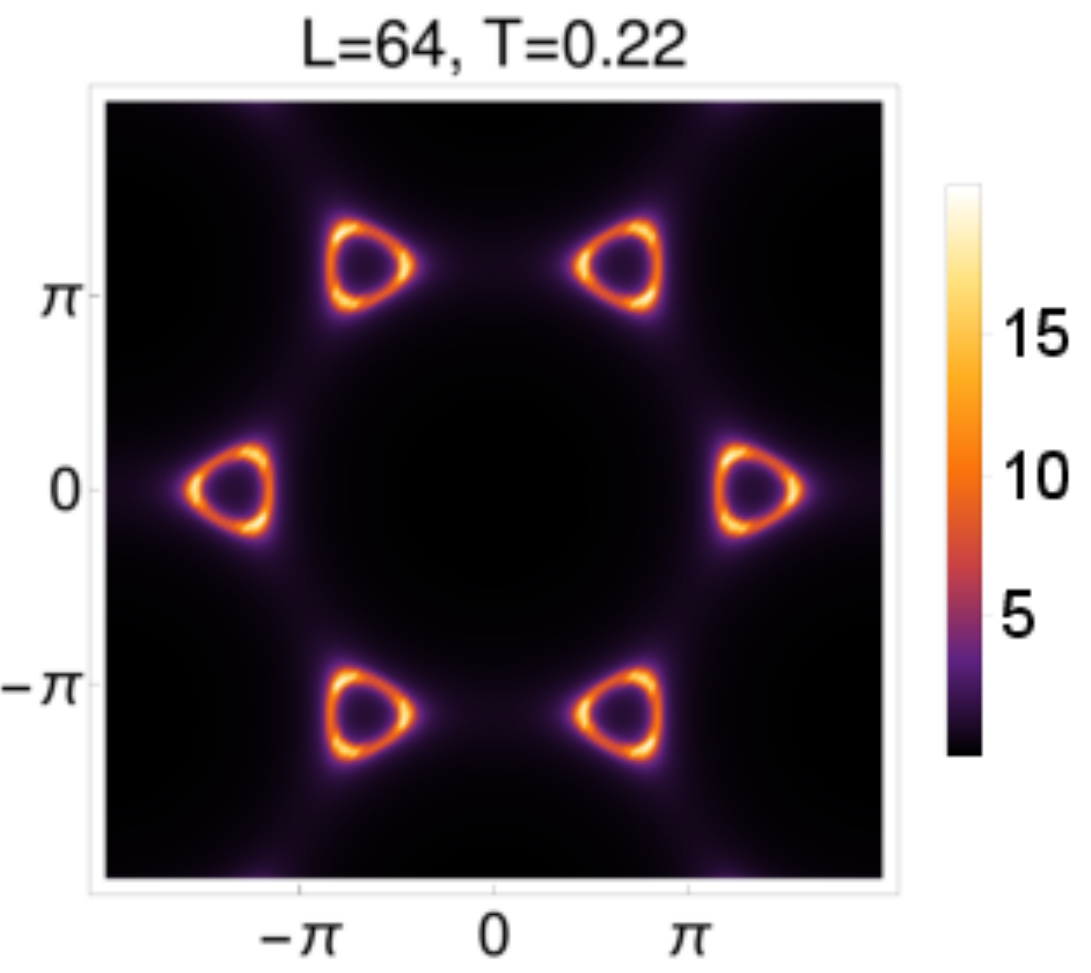}{\label{MCT0.22}}}\\
  \subfloat[T=0.35, L=64]{\includegraphics[width= 0.48\columnwidth]{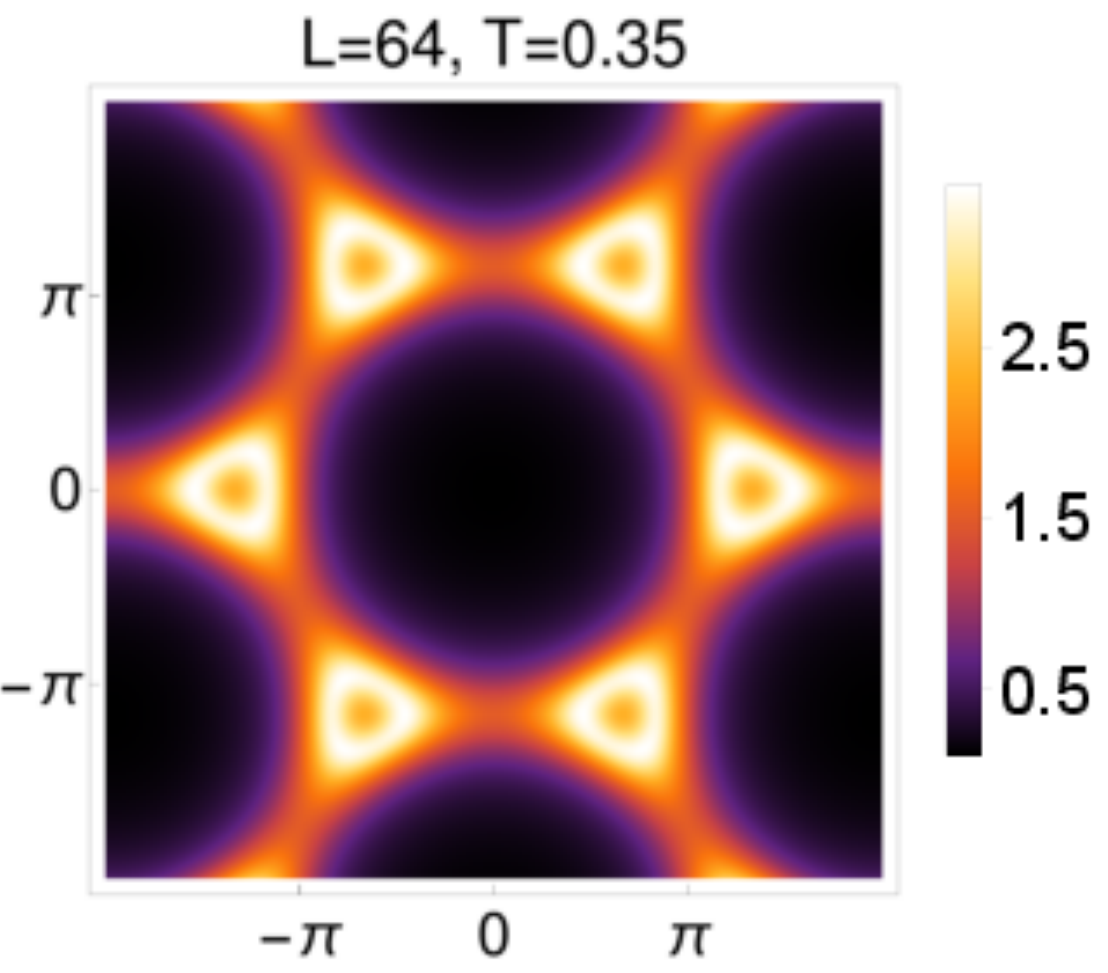}{\label{MCT0.35}}}
  \subfloat[T=3.50, L=64]{\includegraphics[width= 0.48\columnwidth]{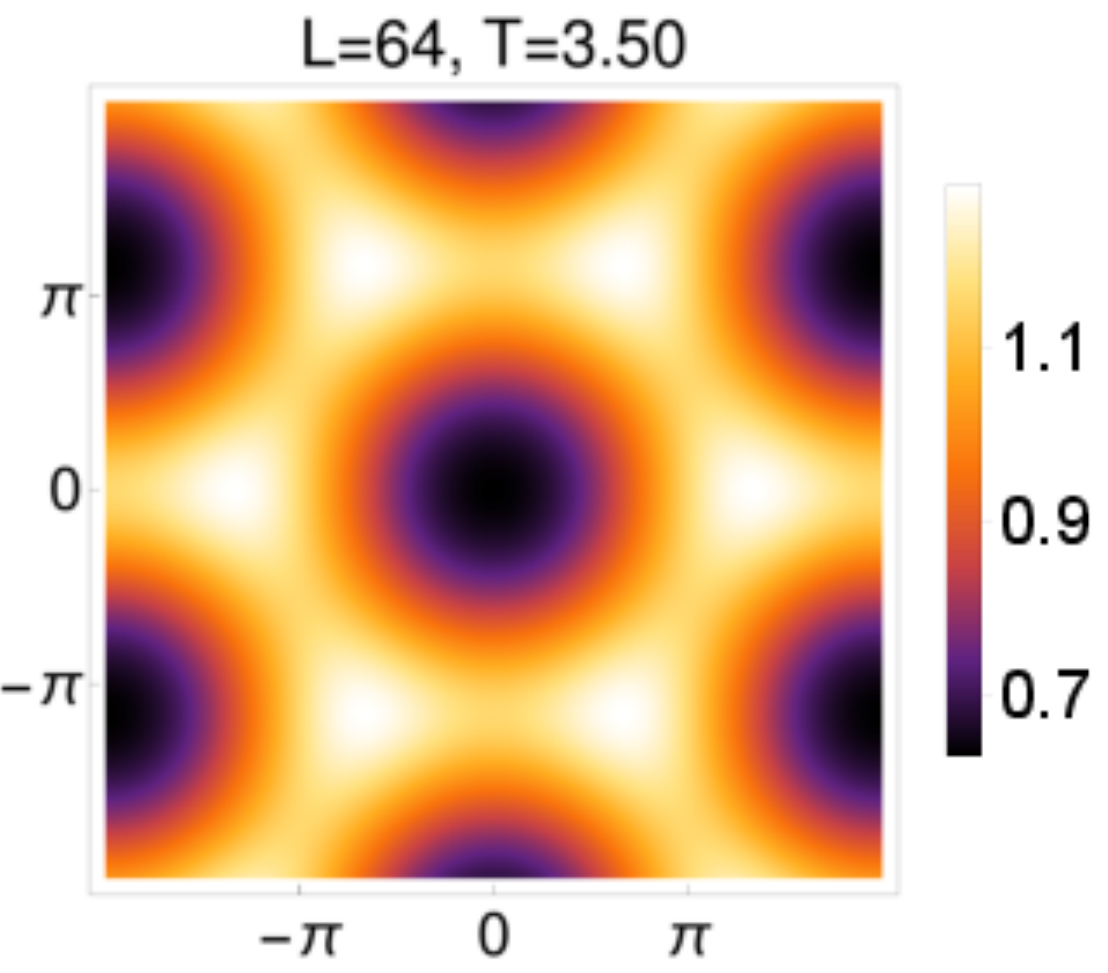}{\label{MCT3.50}}}
  \caption{Temperature and in-plane momentum dependence (with out of plane momentum
set to zero) of the equal time correlation function of spins in the same layer (sublattice), obtained from
    classical Monte Carlo simulations of the effective model for a system of $L \times L$ unit cells with $L=64$. (a) Data at $T=0.20 (J/4)  \simeq 58 \rm{mK}$ shows clear evidence of
    the entropic selection of zone-boundary spiral wavevectors (see main text for details). (b) Data at a slightly higher temperature $T=0.22 (J/4) \simeq 64 \rm{mK}$ shows nearly equal intensity all along the locus of spiral wavevectors favoured by the exchange interactions. (c) This weight
    along the locus of spiral wavevectors is already visible at a slightly higher temperature $T=0.35 (J/4) \simeq 100 \rm{mK}$. (d) Finally,
    at an even higher temperature $T=3.50 \simeq 1 \rm{K}$, the momentum dependence has no sharp features ($J/4 \simeq 290 \rm{mK}$).}
    \label{eqtimecorr}
  \end{figure}
  \begin{figure}[h]
    \centering
    \includegraphics[width=\columnwidth]{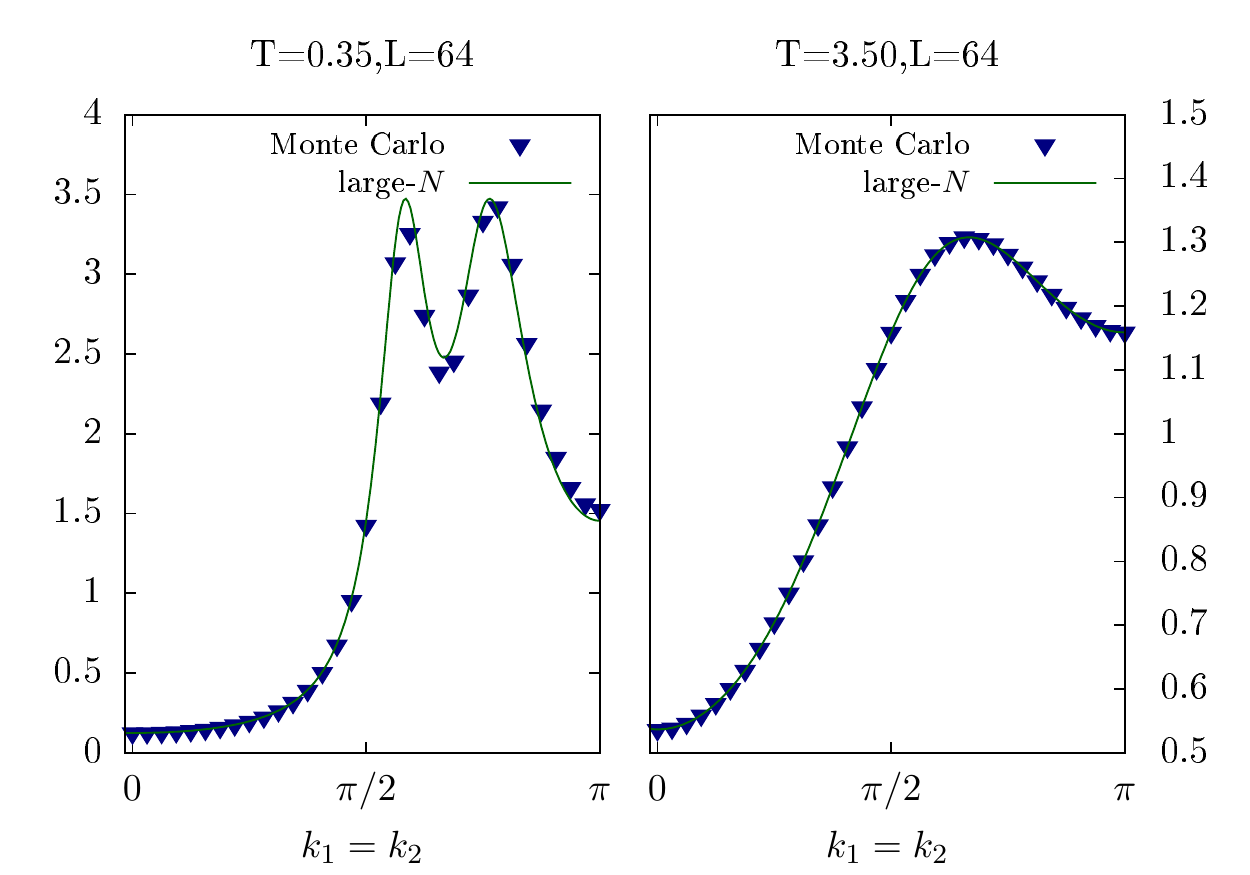}
    \caption{Intra-layer spin correlations in momentum space ($y$ axis) obtained using classical Monte Carlo simulations of the effective model  for a system of $L \times L$ unit cells with $L=64$ are well-approximated by large-$N$ (Sec.~\ref{largen}) results for the same quantity. The correlation functions are plotted along the cut $k_1=k_2$ in momentum space (with out of plane momentum set to zero). The left panel shows this comparison for $T=0.35 (J/4) \simeq 100 \rm{mK}$. The right panel shows the same
    comparison for $T=3.50 (J/4) \simeq 1 \rm{K}$ ($J/4 \simeq 290 \rm{mK}$).}
    \label{comparison}
  \end{figure}
  \begin{figure}[h]
  \centering
  \includegraphics[width=\columnwidth]{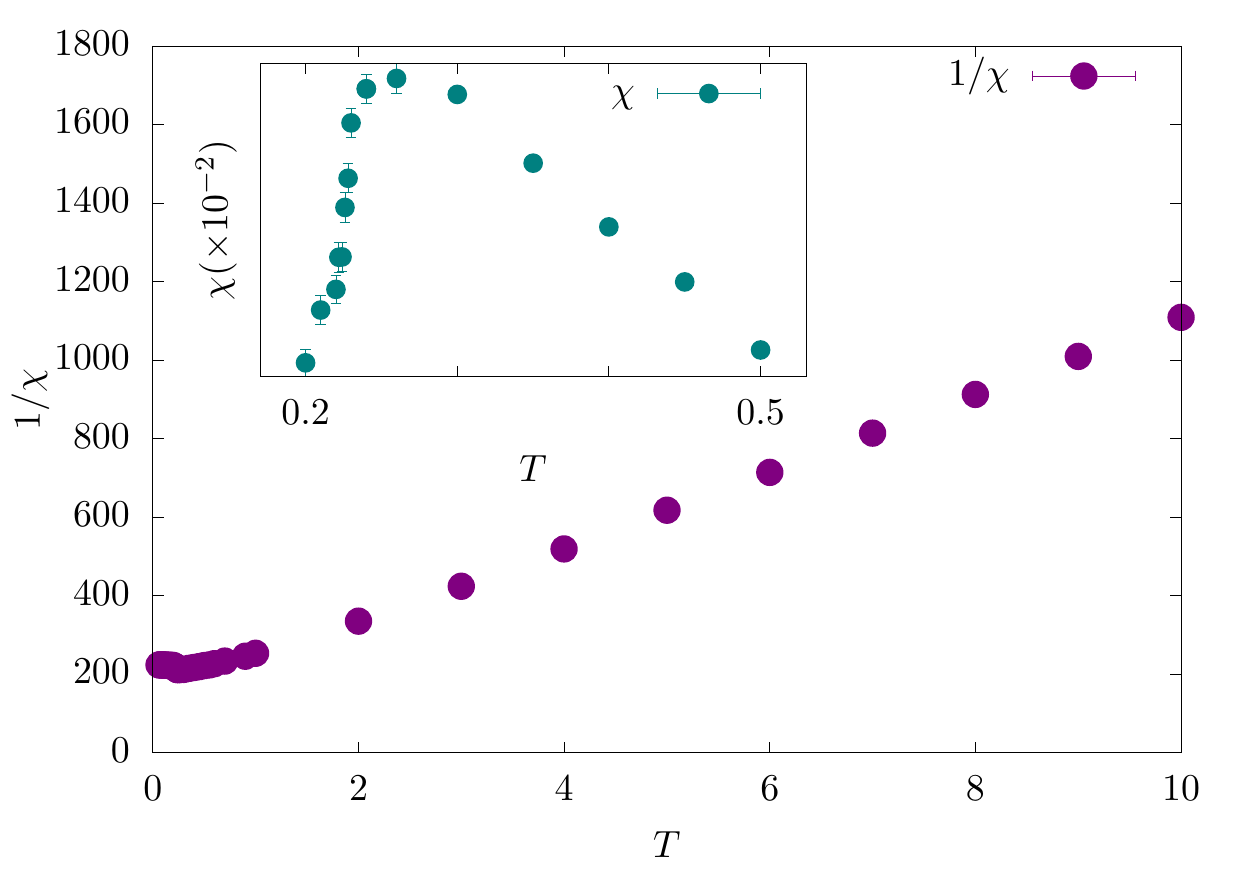}
  \caption{Inverse susceptibility $1/\chi$ of the effective model on a lattice of $L \times L$ unit cells, with $L=64$,
    plotted against temperature $T$ expressed in units of $(J/4) \approx 290 \rm{mK}$ ($\chi$ is defined as in Eq.~\eqref{chidef}). A clear deviation from linearity is visible at low temperature. Inset: The uniform susceptibility $\chi$
    at low temperature shows a crossover at a temperature roughly consistent with the peak in the specific heat data. This crossover temperature corresponds to the temperature scale at which spiral
  correlations start to build up (as evidenced by our results for the spin correlations and structure factor), although our spinwave calculations at $T=0$ strongly suggest that long-range spiral order (favoured at $T=0$ by the pattern of exchange couplings) is destabilized by singular spinwave fluctuations. Note that the crossover scale is
  consistent with the position of the peak in the specific heat curve, which marks
  the sharp onset of nematic order in the bond energies.}
  \label{chi}
  \end{figure}
  \begin{figure}[t]
  \centering
  \includegraphics[width=\columnwidth]{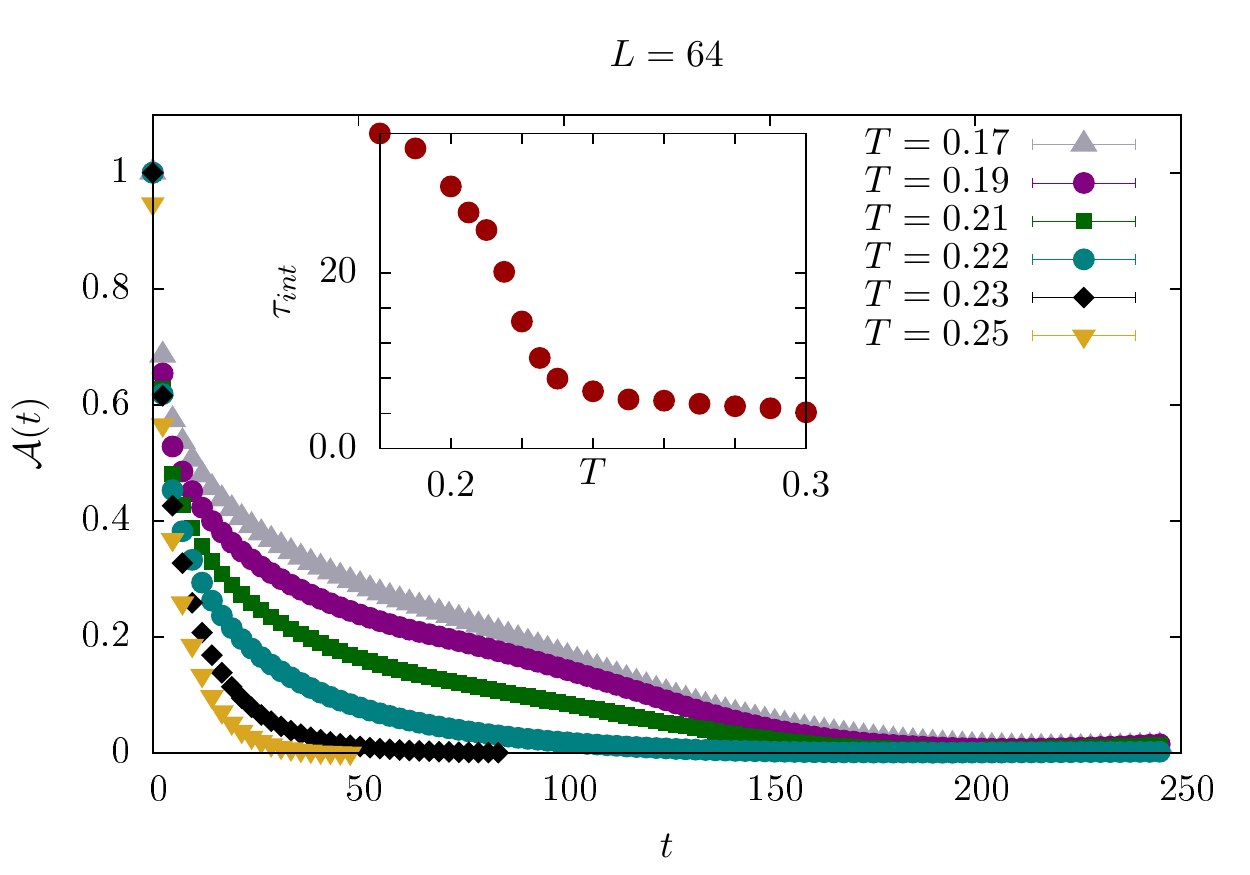}
  \caption{Spin autocorrelation function $\mathcal{A}(t)$ of the effective model plotted as a function of $t$ displayed in units of $(J/4)^{-1}$. Inset shows the temperature dependence of the relaxation time defined via the integrated autocorrelation function. 
  These relaxation times show a crossover at $T_{\rm crossover} \simeq 0.22 (J/4) \simeq 64 \rm{mK}$. The uniform susceptibility display crossovers at roughly the same temperature. As mentioned earlier, this crossover temperature corresponds to the temperature scale at which spiral
  correlations start to build up (as evidenced by our results on the spin correlations and structure factor), although our spinwave calculations at $T=0$ strongly suggest that long-range spiral order (favoured at $T=0$ by the pattern of exchange couplings) is destabilized by singular spinwave fluctuations. Note that the crossover scale is
  consistent with the position of the peak in the specific heat curve, which marks
  the sharp onset of nematic order in the bond energies.}
\label{spincorr}
\end{figure}

  To study the dynamics, we consider the classical Hamiltonian equations of motion, 
  given by
  \begin{equation}
    |S|\frac{d\hat{n}_i}{dt}=\sum_j M_{ij}\hat{n_j} \times \hat{n_i},
    \label{dyn}
  \end{equation}
  where $\hat{n}_i$ are unit-vectors satisfying $\hat{n}_i^{2}=1$. The connectivity matrix $M_{ij}$ is defined in Eq.~\eqref{effec} and given given by the pattern of couplings in Fig.~\ref{effec} with couplings rescaled by a factor of $|S|^2$.
  Following previous work on dynamics of spin models~\cite{Keren,Moessner_Chalker_prl,Moessner_Chalker_prb}, we integrate the Hamiltonian equations of motion numerically using the fourth-order
  Runge-Kutta method. The time step of the numerical integrator is kept low enough
  to ensure the energy remains conserved to within the accuracy needed. In practice, we use a time step of $0.03 (J/4)^{-1}$ to achieve this. To obtain the dynamical correlation functions,
  we integrate the Hamiltonian equations of motion starting from different initial configurations
  generated by the Monte Carlo simulation described in the previous paragraph. All quantities 
  are averaged over initial conditions and the frequency dependence of observables is calculated by averaging Fourier transforms
  of the time evolution of the observable over this ensemble of initial conditions.
  \begin{figure}[h]
    \centering
    \includegraphics[width=\columnwidth]{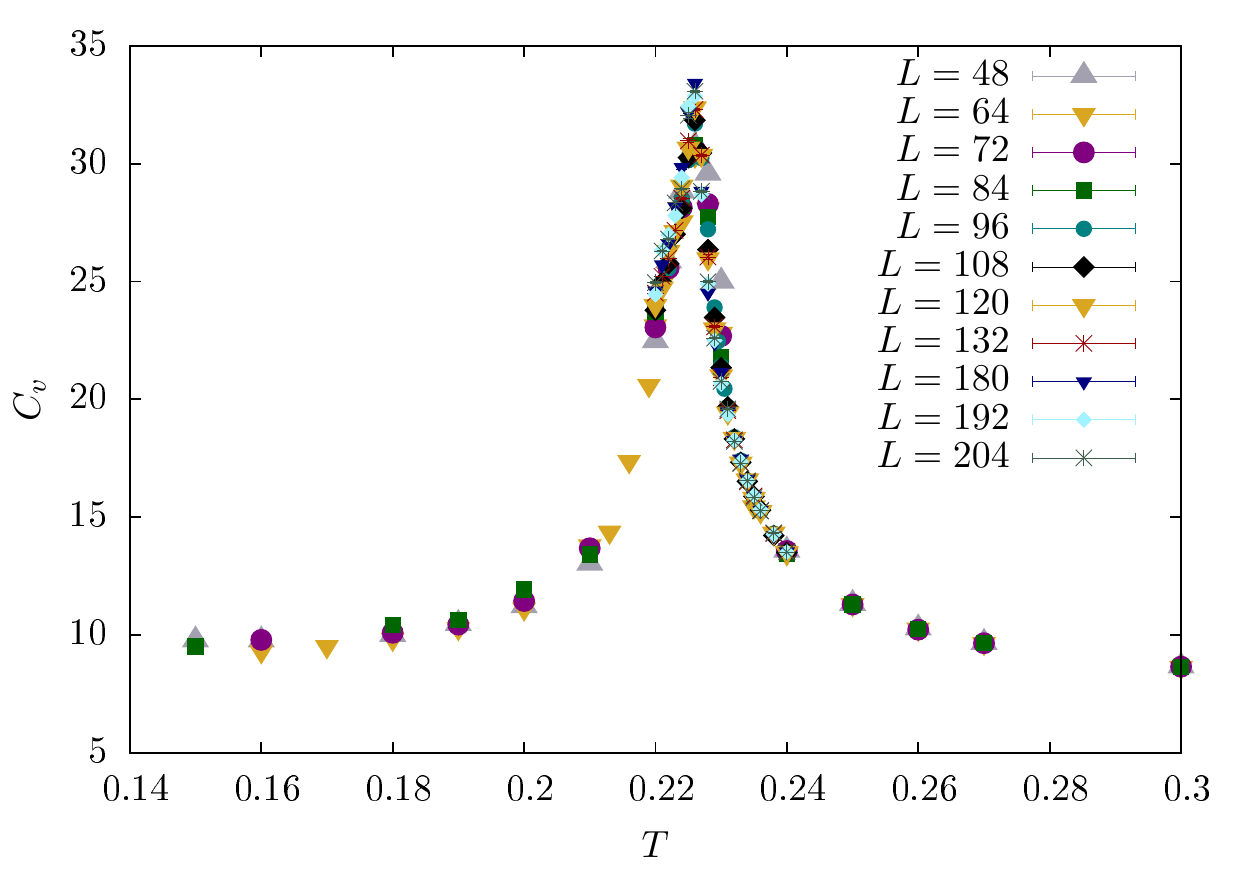}
    \caption{Specific heat $C$ (Eq.~\eqref{cv}) of the effective model on a lattice of $L \times L$ unit cells, with $L=48$, $64$, $72$, $84$, $96$, $108$, $120$, $132$, $180$, $192$  and 
    $204$, plotted against temperature $T$, expressed in units of $(J/4) \approx 290$ mK. There is a clear peak at
      a temperature $T^{*} \simeq 0.22 (J/4) \simeq 64 \rm{mK}$. This peak does not scale with system size, apparently ruling out a phase transition. Indeed,
      our results appear to saturate to the thermodynamic limit already for the range
      of sizes studied, including at the position of the peak. However, results
      for the bond-energy nematic order parameter for the same range of sizes suggest the sharp onset of nematicity
      at a temperature corresponding to this peak (see below). }
  \label{specH}
  \end{figure}

   \begin{figure}[h]
    \centering
    \includegraphics[width=\columnwidth]{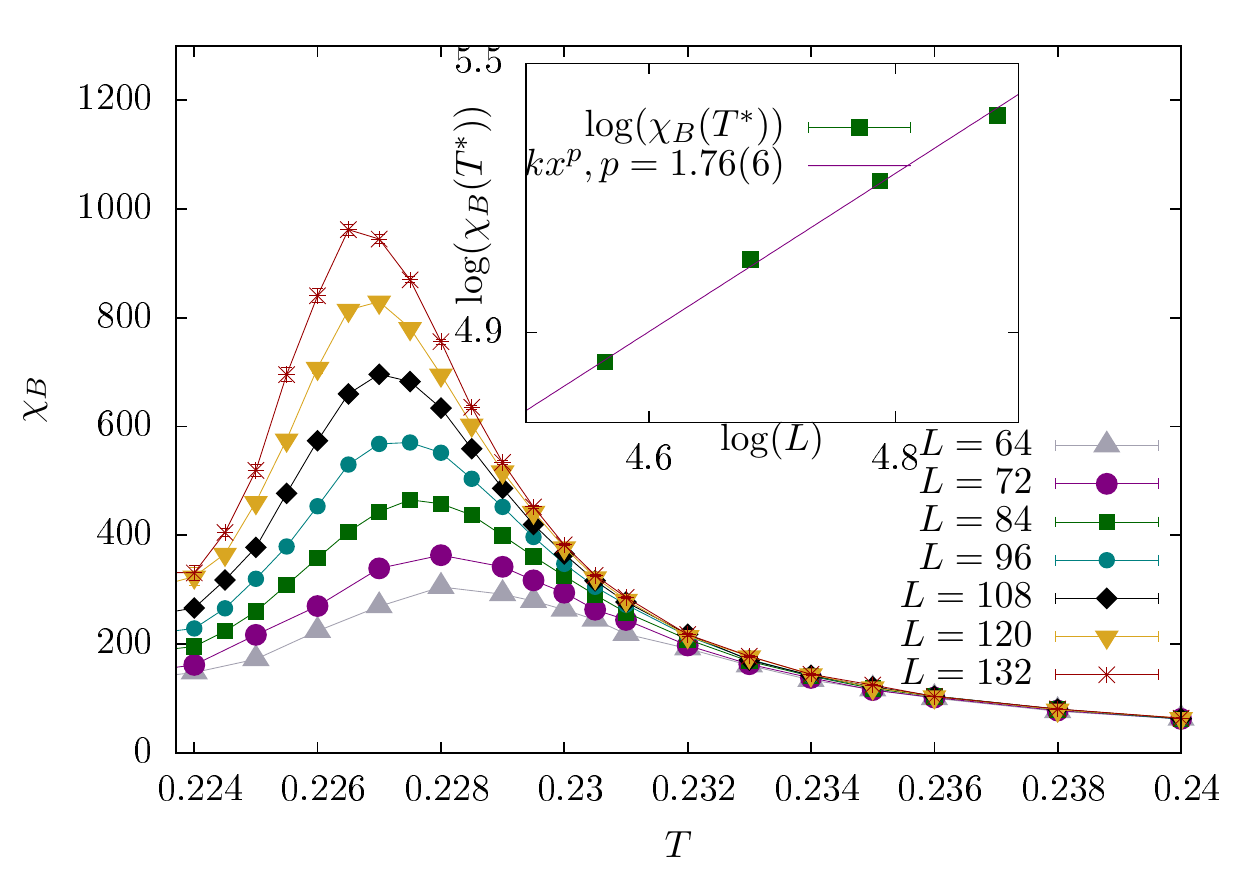}
    \caption{Nematic order parameter susceptibility (Eq.~\eqref{nemchidef}) of the effective model on a lattice of $L \times L$ unit cells, with $64$, $72$, $84$, $96$, $108$, $120$ and $132$,
    plotted against temperature $T$, expressed in units of $(J/4) \approx 290$ mK. There is a clear peak at
      a temperature $T^{*} \simeq 0.22 (J/4) \simeq 64 \rm{mK}$. The height of this peak, plotted in the inset as a function of system size,  shows
      the expected finite-size scaling behaviour at a thermodynamic phase transition, consistent with the results of Mulder {\em et. al.}\cite{Arun}. In particular, our power-law fit (shown as a line in the inset) for the $L$ dependence of the peak height has power-law exponent $1.76(6)$, consistent with the known value of $26/15 = 1.733\dots$ for this exponent at the three-state Potts transition.}
  \label{orderparameter}
  \end{figure}

\begin{figure}[h]
  \centering
  \subfloat[]{\includegraphics[width= 0.48\columnwidth]{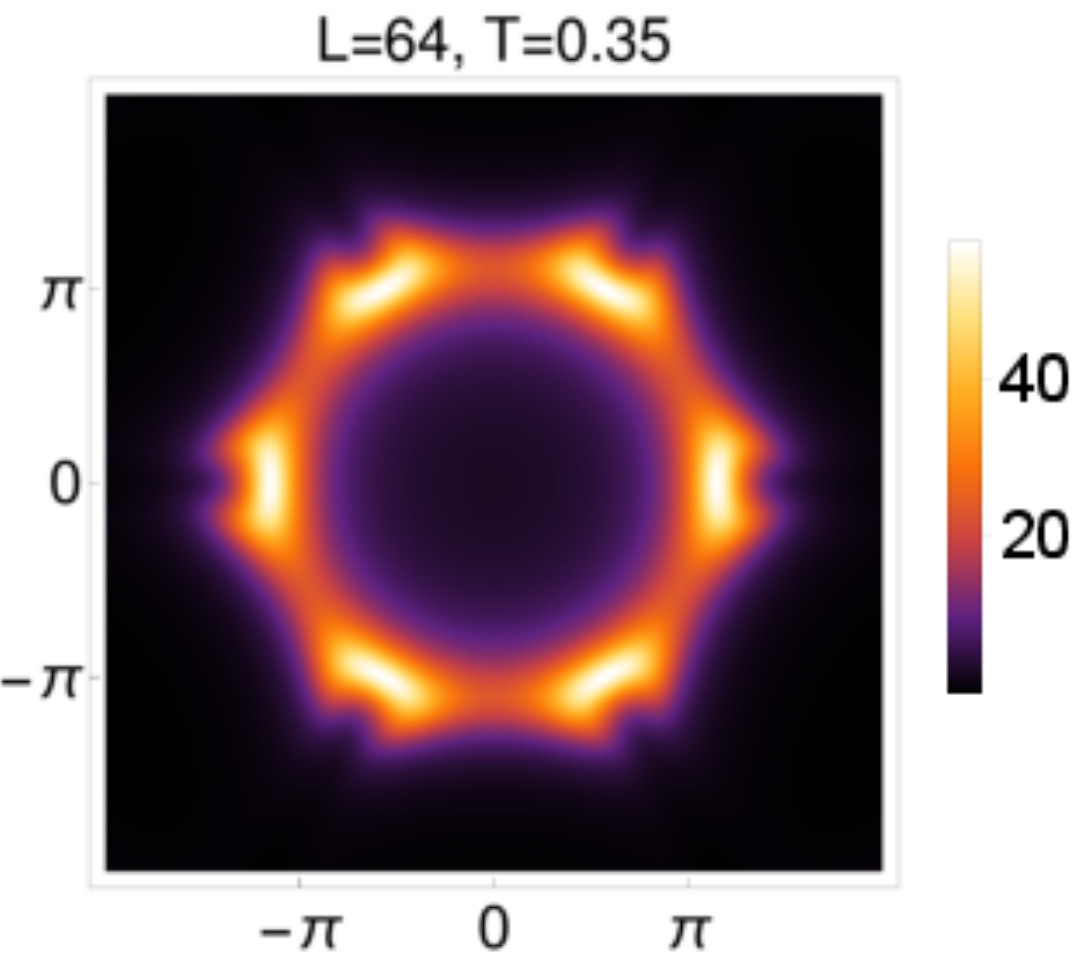}}{\label{mcsfac1}}
  \subfloat[]{\includegraphics[width= 0.48\columnwidth]{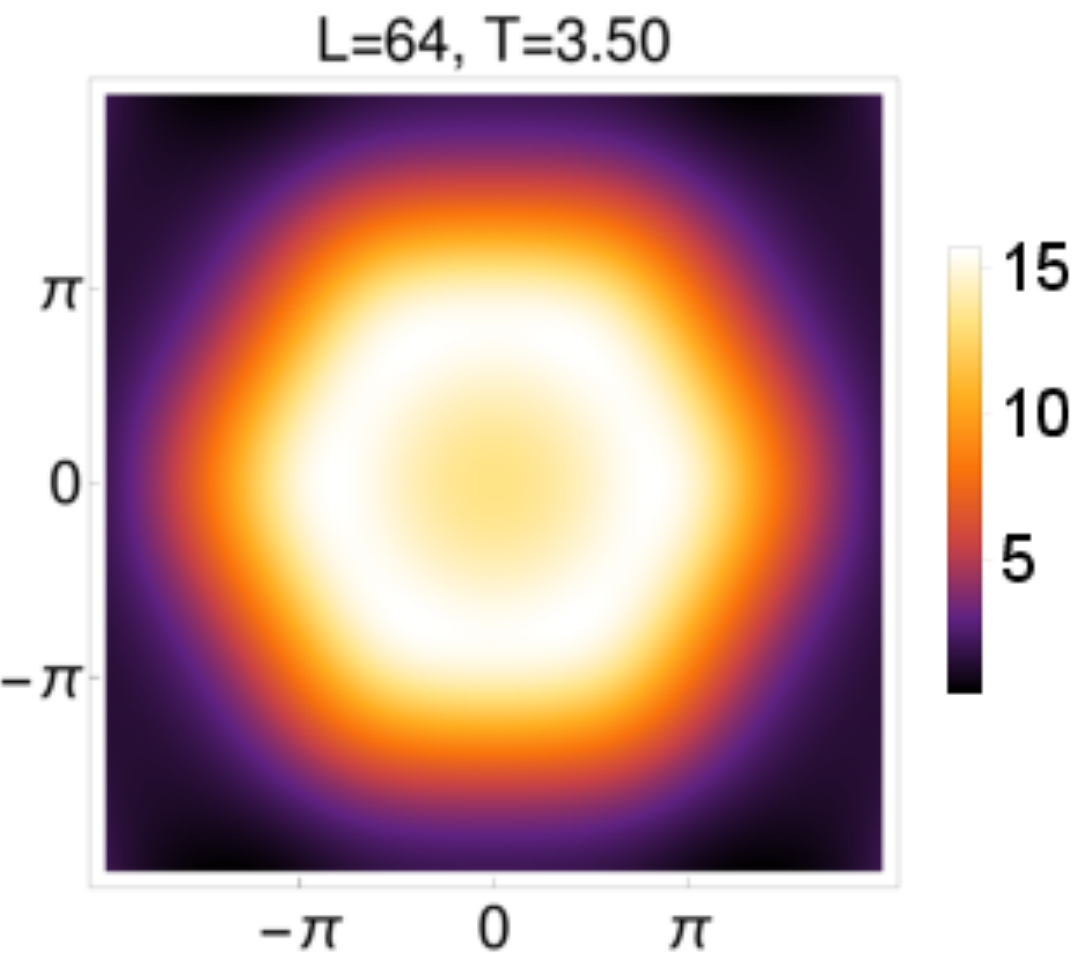}}{\label{mcsfac2}}
  \caption{In plane momentum dependence (with out of plane momentum set to zero) of the equal-time structure factor $\mathcal{S}^{\rm MC}(\vec{k})$ (Eq.~\eqref{mcsfaceq}) of the effective model, obtained from
  Monte Carlo simulations of systems with $L \times L$ unit cells, with $L=64$,  for temperatures (a) $0.35 (J/4) \simeq 100 \rm{mK}$ and 
  (b) $3.50 (J/4) \simeq 1 \rm{K}$ ($J/4 \simeq 290 \rm{mK}$). The spiral features visible in the corresponding
  intra-layer correlation function at the lower temperature (displayed earlier in Fig.~{\protect{\ref{eqtimecorr}}}) are partially smeared out due to the effect of form factors, but still visible. The results at the higher temperature are largely featureless.}
  \label{mcsfac}
\end{figure}

\subsection{Results}
\label{results}
The Mermin-Wagner theorem rules
out the spontaneous breaking of any continuous symmetry in two dimensions, thereby
ruling out any nonzero temperature regime with true long range spiral order
in the spin correlations. However, discrete lattice symmetries can 
still be broken. Indeed, the work of Mulder {\em et. al.}\cite{Arun} has demonstrated
an apparent transition to bond-energy nematic order for $J_2/J_1 =-1$ (and nearby values) in our notation, {\em i.e.} with both couplings antiferromagnetic (as far as the classical physics is concerned, the sign of $J_1$ can
be changed by flipping the spins on one sublattice, connecting this result to the case
of interest to us). A similar transition had also been reported earlier in the literature by Okumura {\em et. al.}\cite{Okumura} for $J_2/J_1>-1/2$ (in our notation). While this transition was seen to be accompanied by the expected singular behaviour
of the specific heat in the cases studied by Okumura {\em et. al.}, Mulder {\em et. al.}'s results
suggested that the specific heat does not scale at the nematic transition
in the regime of $J_2/J_1$ studied by them.\cite{Arun,Okumura}

From the point of view of the experiments that form our motivation, it is important
to ask what are the signatures in the spin channel of this puzzling onset of bond-energy nematicity at $J_2/J_1=1$? To address this question, we study the effective model on triangular lattices with $L \times L$ unit cells, with each unit cell
having  two basis spins, and obtain the spin correlators, uniform spin susceptibility
and the local spin autocorrelation function in the low temperature regime.

First, we look at the Fourier transformed correlation function of spins in the same plane 
$\Big<n_{\alpha}(\vec{k})n_{\alpha}(-\vec{k})\Big>_{\rm MC}$,
obtained easily in our Monte Carlo simulations by fast Fourier transforming the spin configurations. At low temperature
below a crossover scale $T_{\rm crossover} \simeq 0.22 (J/4) \simeq 64 \rm{mK}$, we see clear evidence for slowly decaying
spiral correlations at wavevectors that form a one-dimensional locus in $q$ space. In fact, this tendency becomes gradually
visible starting at somewhat higher temperatures. When the temperature is lowered below this crossover scale, order-by-disorder
effects apparently start preferring a particular set of zone boundary spiral wavevectors from this locus of degenerate spirals (Fig.~\ref{MCT0.20}). This is
consistent with the behaviour expected from the classical analysis of fluctuations about these spiral states in Sec.~\ref{class_sel}, since it is the same set of wavevectors that is selected. The full locus of spiral wave vectors (Eq.~\eqref{Minima} and Fig.~\ref{Minima}) 
obtained from large-N calculations in Sec.~\ref{largen} become visible at somewhat higher temperature, 
as shown in Fig.~\ref{MCT0.22}. At even higher temperatures, the 
correlation function between spins in the same layer starts looking more and more liquid-like, as shown in Fig.~\ref{MCT0.35} and Fig.~\ref{MCT3.50}. Further, the correlation functions obtained in the Monte-Carlo simulations are in reasonable agreement with the ones calculated in large-$N$. We have displayed the agreement of our Monte-Carlo correlation functions within the same layer with the large-$N$ results in Fig.~\ref{comparison}. The slight disagreement at the lower temperature can be ascribed to the fact that the large-$N$ analysis does not capture the entropic effects which lead to the selection of a particular set of spiral wave-vectors at low temperatures, as described in Sec.~\ref{class_sel}. 

Next we compute the uniform susceptibility, $\chi$, given by 
\begin{equation}
  \chi=\frac{1}{2T L^2}\Big(\langle \sum_{i}\hat{n}_{i}^2\rangle_{\rm{MC}}
  -\langle  \sum_{i}\hat{n}_{i}\rangle^{2}_{\rm{MC}} \Big).
  \label{chidef}
\end{equation}.
In Fig.~\ref{chi}.we display results for the inverse spin susceptibility, $1/\chi$.
The linear behaviour at high temperature, characteristic of a paramagnet, persists down to a crossover temperature, below which
deviations are apparent. [The linear behavior, if extrapolated down, has an antiferromagnetic intercept, which reflects the fact that we are working with an effective model of $S=3/2$ spins, and the true high-temperature limit (at temperatures well above the large ferromagnetic exchange couplings) is not accessible to our model.] Deviations from paramagnetic behaviour below the crossover scale are also apparent in the plot of the uniform susceptibility $\chi$ shown in the inset of Fig.~\ref{chi}. 
Note that the small bump in $\chi$ as a function of temperature serves as a marker
for the crossover temperature, which is consistent with the crossover visible in the Fourier transform of the spin correlators discussed earlier.

Next we we look at spin autocorrelation functions, defined as 
\begin{equation}
  \mathcal{A}(t)= \langle \hat{n}_{i}(0) \cdot  \hat{n}_{i}(t) \rangle_{\rm{MC}}
\end{equation}
We show the decay of spin autocorrelations in Fig.~\ref{spincorr}. At higher temperatures,
the autocorrelations decay exponentially like in a paramagnet. At lower temperatures,
the autocorrelation curves develop a knee and cross over to a regime of slow dynamics. To extract a time-scale
from these relaxation rates, we define the integrated autocorrelation time $\tau_{\rm{int}}$ as
\begin{equation}
  \tau_{\rm{int}}=\int_{0}^{\infty}dt \mathcal{A}(t)
\end{equation}
We plot the relaxation time scales $\tau_{\rm{int}}$ obtained in this manner in the inset of Fig.~\ref{spincorr}. We see that
the autocorrelation timescale shows a crossover to slow dynamics at $T_{\rm{crossover}} \simeq 0.22 (J/4)$, consistent with the crossover in the uniform susceptibility plots
and the Fourier transform of the spin correlation functions.

To connect this crossover in the spin channel with the puzzling transition to nematic order in the bond-energies
reported earlier in Mulder {\em et. al.}\cite{Arun} for $J_2/J_1=1$ and nearby values, we have
revisited the specific heat and nematic order parameter suscepbility of this system, going
to somewhat larger sizes than in the work of Mulder {\em et. al}.
Defining the specific heat as
\begin{equation}
  C= \frac{1}{2T^2 L^2} \Big(\langle E^2 \rangle_{\rm{MC}}-\langle E \rangle^2_{MC}\Big),
  \label{cv}
\end{equation}
where $E$ is the total energy of a configuration and $\langle \dots\rangle_{\rm{MCS}}$ denote a Monte Carlo average, we have
obtained the specific heat data for different system sizes shown in Fig.~\ref{specH}.
We see a peak in the specific heat at $T \approx 0.22 (J/4)$. However, we also note that
the peak does not
scale at all with the system size. Indeed,  from Fig.~\ref{specH}, we see that linear sizes that differ by more than a factor of three give curves that overlap with each other within error bars,
indicating that finite size effects are already negligible at these sizes.  Note that this peak
is apparently unrelated to the bump at $T \simeq 4 \rm{K}$ in the experimental specific heat curve reported in Ref.~\onlinecite{Balz}: 
Indeed this temperature scale seen in the experiments corresponds quite well to the average of the two energy scales (since the ferromagnetic couplings in the two layers are different) associated with the unbinding of the ferromagnetically bound effective $S=3/2$ moments into three $S=1/2$ moments , suggesting that this is the origin of the specific heat feature studied experimentally.
Since our calculations are in terms of an effective Hamiltonian for the spin $S=3/2$ degrees of freedom, we do not
capture this higher temperature feature within our effective theory.

Turning our attention to the interpretation of the peak in the specific heat at $T \approx 0.22 (J/4)$, we note that any interpretation of
this specific heat peak in terms of a thermodynamic singularity associated with a phase
transition would normally have been ruled out by the fact that the data appears
to have already converged to the thermodynamic limit over the range of sizes studied.
(We have also checked that the spin structure factor data (discussed below) 
and the equal time correlation results (displayed earlier) for spins are both reasonably well-converged to the thermodynamic limit at the sizes used in our study, suggesting
that this range of sizes is perfectly adequate as a means of extrapolating to the
thermodynamic limit.)

However, as was already noted by Mulder {\em et. al.},\cite{Arun} when one computes for the same range of sizes the complex bond-energy nematic order parameter defined as:
\begin{equation}
  B(\vec{r})=\hat{n}_{1,\vec{r}}\cdot \hat{n}_{2,\vec{r}} +e^{i 2\pi/3}\hat{n}_{1,\vec{r}}\cdot \hat{n}_{2,\vec{r}+\hat{e}_1} +e^{i 4\pi/3}\hat{n}_{1,\vec{r}}\cdot \hat{n}_{2,\vec{r}+\hat{e}_2},
  \label{nematic}
\end{equation}
we see behaviour that is consistent with the sharp onset of nematicity at a temperature
corresponding to this peak in the specific heat. This is shown in Fig.~\ref{orderparameter} where
we  plot the order parameter susceptibility $\chi_B$, given by
\begin{equation}
  \chi_B=\frac{1}{T L^2}\Big(\langle \sum_{\vec{r}}|B(\vec{r})|^2\rangle_{\rm{MC}}
  -\langle  \sum_{\vec{r}}B(\vec{r})\rangle^{2}_{\rm{MC}} \Big),
  \label{nemchidef}
\end{equation}
over a somewhat larger
range of sizes than in the previous work.\cite{Arun} Clearly, we see behaviour consistent with Mulder {\em et. al.}'s identification of a transition
to nematic order in the bond-energies.\cite{Arun}. In particular, we are able
to fit the height of the peak to the expected scaling behaviour at the three-state
Potts transition (inset of  Fig.~\ref{orderparameter}).
More work is needed to understand this puzzling nematic transition, given
that the discrepancy between the behaviour of the specific heat and the order parameter
susceptibility is seen to persist even at the larger sizes accessed in our study.

Independent of this puzzle, we can nevertheless conclude that the temperature of specific heat peak is roughly consistent with the crossover in the spin channel associated with a growing spiral correlation
length (Fig.~\ref{eqtimecorr}) which leaves its mark on the Fourier transform of
the spin correlation function, on the uniform spin susceptibility, and on the local
spin autocorrelation function.
 
In addition, we have also measured the equal time structure factors defined as:
\begin{equation}
  \mathcal{S}^{\rm{MC}}(\vec{k})=\frac{1}{L^2}\langle \lvert \hat{n}_{1}( \vec{k})f_{1}(\vec{k}) + \hat{n}_{2}(\vec{k})f_{2}(\vec{k}) \rvert^{2} \rangle_{\rm{MC}},
  \label{mcsfaceq}
\end{equation}
where the form factors of the effective $S=3/2$ moments are given in Appendix~\ref{formfacs}. The equal time structure factors for $T=0.35 (J/4) (\approx 100 mK)$ and $T=3.50 (J/4) (\approx 1 K)$ are shown in Fig.~\ref{mcsfac}. At the lower temperature, we see clear evidence of spiral correlations, whereas
the higher temperature results are featureless.

\begin{figure}[h]
  \centering
  \subfloat[]{\includegraphics[width= 0.48\columnwidth]{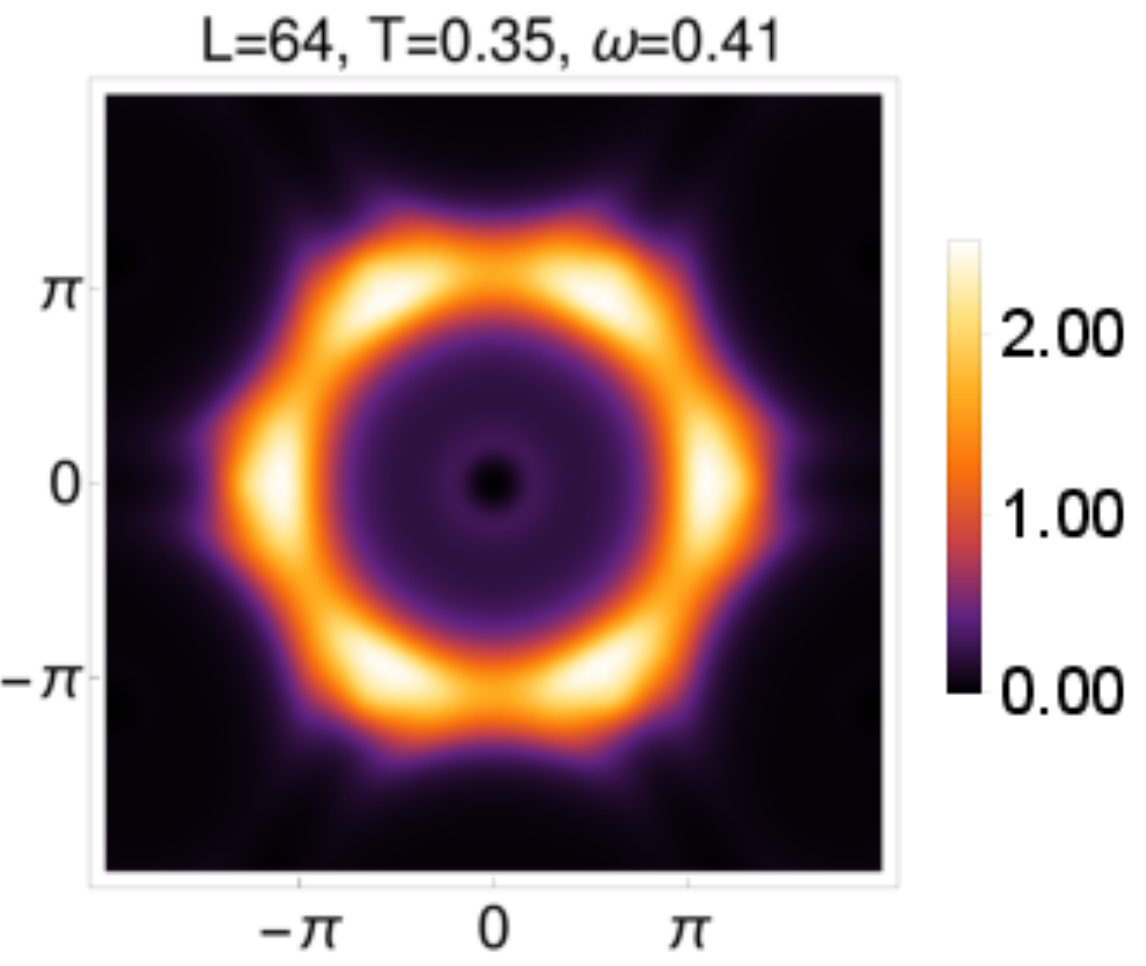}}{\label{dynsfac1}}
  \subfloat[]{\includegraphics[width= 0.48\columnwidth]{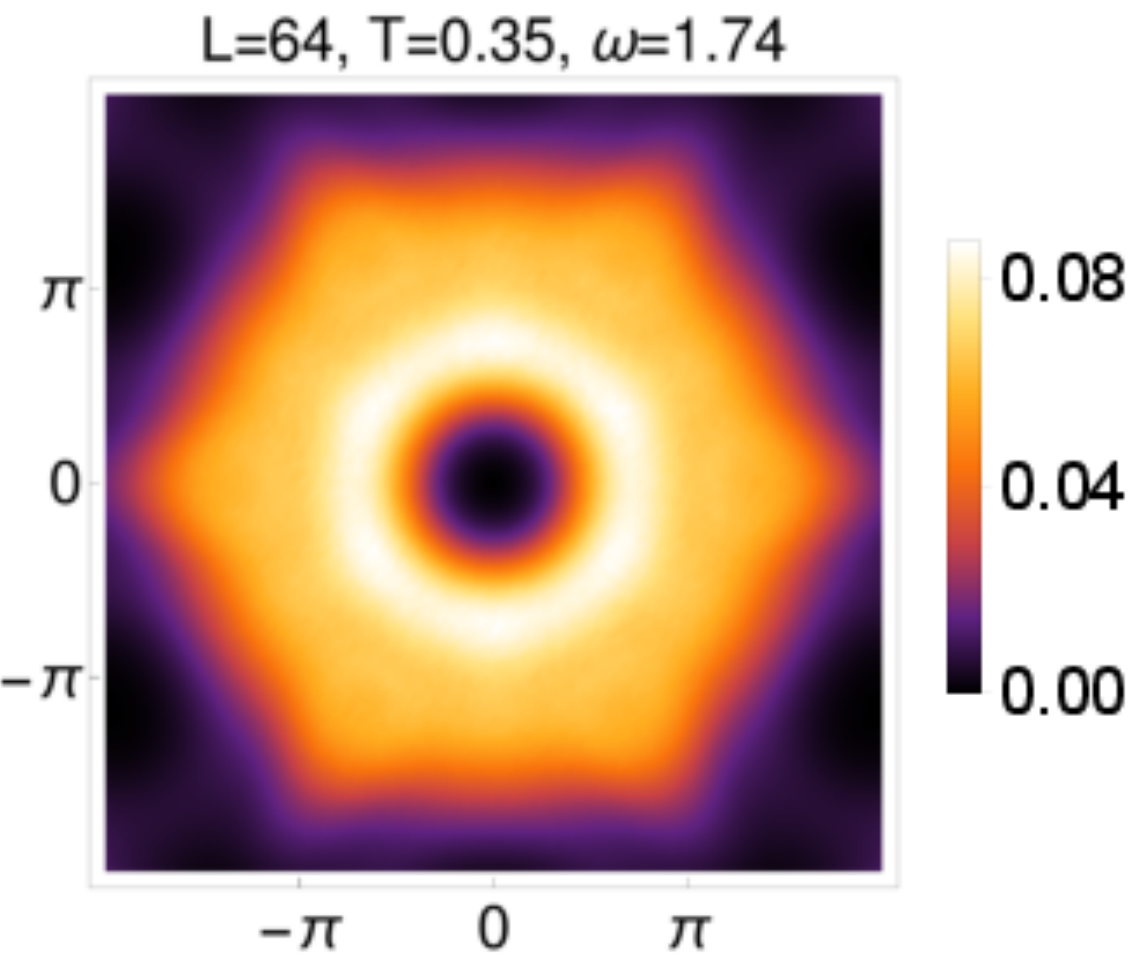}}{\label{dynsfac2}}
  \caption{Plots showing the in plane momentum dependence (with out of plane momentum set to zero) of dynamic structure factors $\mathcal{S}^{\rm MC}(\vec{k},\omega)$ defined in Eq.~\eqref{mcdsfac}, obtained from combined Monte Carlo-Molecular Dynamics simulations of the effective at low temperature $T= 0.35(J/4) \simeq 100$ mK for a 
  system of $L \times L$ unit cells with $L=64$. a) The dynamic structure factor at low frequency ($\omega=0.41 (J/4) \simeq 0.01 \rm{meV}$ shows clear features corresponding to low-frequency fluctuations at wavevectors on the spiral locus. (b) The dynamic structure factor at a somewhat higher frequency
  $\omega=1.74 (J/4) \simeq 0.0435 \rm{meV}$ (which is still very low compared to the scale at which inelastic neutron scattering experiments
  have probed the dynamics) is already featureless ($J/4 \approx 290$ mK). Note that recent experiments have probed the dynamic structure factor mainly at significantly higher frequencies ($\gtrsim 0.25 \rm{meV}$), which actually correspond in our picture to the natural energy scale for transitions of the strongly coupled
ferromagnetic triangles from the total spin $S=3/2$ multiplet to the higher energy $S=1/2$ doublets.} 
\label{dynsfac}
\end{figure}

Finally, we have calculated the dynamic structure factor, defined as
\begin{equation}
  \mathcal{S}^{\rm{MC}}(\vec{k},\omega)=\frac{1}{L^2 N_{\tau}}\langle \lvert \hat{n}_{1}( \vec{k},\omega)f_{1}(\vec{k}) + 
  \hat{n}_{2}(\vec{k},\omega)f_{2}(\vec{k}) \rvert^{2} \rangle_{\rm{MC}}.
  \label{mcdsfac}
\end{equation} 
Here, $\hat{n}_{\alpha}(\vec{k},\omega)$ is calculated
by fast Fourier transforming $\hat{n}_{\alpha,}(r_i,t)$ obtained from the numerical integration of the Hamiltonian
equations of motion ~\eqref{dyn} [$N_{\tau}$ is the number of steps used in numerical integration].
The dynamic structure factors at $T=0.35 (J/4)\approx 100 \rm{mK}$ for frequencies $0.41 (J/4) \approx 0.010 $ \rm{meV}
and $1.74 (J/4)\approx 0.044 $ \rm{meV} are shown in Fig.~\ref{dynsfac}.  Both these frequencies fall well-within the quasi-elastic
window of the recent inelastic neutron scattering measurements.\cite{Balz} At the lower of the two frequencies, one sees clear
features corresponding  to low-frequency fluctuations at wavevectors on the spiral locus. At the higher frequency, the structure
factor is liquid-like and relatively featureless.

\section{Discussion}
\label{discussion}
The analysis presented here strongly suggests that the low temperature behaviour of $\mathrm{Ca}_{10}\mathrm{Cr}_{7}\mathrm{O}_{28}$ provides
an interesting example of a frustrated magnet in which the exchange couplings favour $T=0$ incommensurate spiral order. The presence of singular spinwave fluctuations at wavevectors in the vicinity of the locus of spiral wavevectors also suggests that spiral order
is unstable at $T=0$ due to these fluctuations, although this leading order spinwave result
itself could get modified by a non-perturbative treatment of $1/S$ corrections. Independent of the fate of the system at $T=0$, we show that there is a nonzero temperature crossover
to a regime in which the spin autocorrelation time-scale, equal time spin correlations, and the dynamic
spin structure factor all reflect the presence of a large but finite correlation length
for spiral spin correlations at a particular set of entropically selected zone boundary spiral wavevectors. The temperature scale for this crossover is roughly the same
as the onset temperature for nematicity in the bond-energies, seen in earlier work.\cite{Arun}

Our numerical
results suggest that this crossover temperature is $T_{\rm crossover} \simeq 0.22 (J/4) \simeq 64 \rm{mK}$---this is small because it is set by the relatively weak effective interactions
between the effective $S=3/2$ degrees of freedom. The
corresponding frequency scale (at which dynamical fluctuations at spiral wavevectors become apparent) is $\omega_{\rm crossover} \simeq 0.4 (J/4) \simeq 0.01 \rm{mev}$, which falls well within the ``quasi-static'' window of inelastic neutron scattering studies of $\mathrm{Ca}_{10}\mathrm{Cr}_{7}\mathrm{O}_{28}$.~\cite{Balz} These recent experiments  have also largely focused on the physics in a somewhat higher temperature window ($T \gtrsim 100 \rm{mK}$)
which is, by our reckoning, significantly above the crossover temperature at which the buildup of spiral correlations could be seen.
In this higher temperature window, our results are quite consistent with the liquid-like behaviour seen in the experiments. In this context, we emphasize
that our analysis, which focuses on the physics of the low-energy effective theory, cannot address the physics of the higher temperature
crossover, corresponding to the ``binding'' of the ferromagnetic triangles into the $S=3/2$ effective moments that form the basic
degrees of freedom at lower temperatures. From a comparison of the relevant energy scales, it appears that at least some
of the features seen in the recent inelastic neutron scattering data on $\mathrm{Ca}_{10}\mathrm{Cr}_{7}\mathrm{O}_{28}$ may be
ascribed to the physics of transitions from the low energy $S=3/2$ multiplet  to higher energy doublets in the
spectrum of the ferromagnetically coupled triangles in each layer. We hope our results provide some stimulus for future
experiments that explore the physics of the crossover to the low temperature regime dominated by the onset of
spiral correlations. After completion of our work, we became aware of a parallel study\cite{NicShannon} that also addresses the physics of $\mathrm{Ca}_{10}\mathrm{Cr}_{7}\mathrm{O}_{28}$, and it would be interesting to compare and contrast our conclusions with those of this parallel study.

\section{Acknowledgements} We thank ICTS-TIFR Bengaluru for hospitality during the School on Current Frontiers
in Condensed Matter Research 2016, where the intial part of this study was completed.
Our subsequent computational work at the TIFR was made possible by the computational resources
of the Department of Theoretical Physics of the Tata Institute of Fundamental
Research, as well as by computational resources funded by 
DST (India) grant  DST-SR/S2/RJN-25/2006. We gratefully acknowledge the hospitality of the Institute of Solid State Physics
at the Univeristy of Tokyo (Kashiwa), where this draft was finalized for submission.

\appendix 
\section{Details of spin-wave calculation}
\label{spinw}

The expressions for the matrix $\mathbf{M}(\vec{q},\vec{k})$ and  $a(\vec{q},\vec{k})$ in Eq.~\eqref{hpbosons} are given here. In the rest of this section, we suppress all explicit $\vec{q}$ dependences in our notation for convenience. Thus we write
\begin{equation}
  \mathbf{M}(\vec{q},\vec{k})= 
  \left(
  \begin{array}{cc}
    \mathbf{A}(\vec{k})  & \mathbf{B}(\vec{k}) \\
    \mathbf{B}(\vec{k}) &  \mathbf{A}(\vec{k}) \\
  \end{array}
  \right)
  \\
\end{equation}
The $2 \times2$ matrices $\mathbf{A}$and $\mathbf{B}$ are given by 

\begin{align}
  \mathbf{A}(\vec{k})=J^{\rm eff}
  \left(
  \begin{array}{cc}
    a(\vec{k})  & c^{*}(\vec{k}) \\
    c(\vec{k}) &  a(\vec{k}) \\
  \end{array}
  \right) 
  \\
  \mathbf{B}(\vec{k})=J^{\rm eff} 
  \left(
  \begin{array}{cc}
    b(\vec{k})  & d^{*}(\vec{k}) \\
    d(\vec{k}) &  b(\vec{k}) \\
  \end{array}
  \right) 
\end{align}
The matrix elements are given by
\begin{align*}
  a(\vec{k})= & (\cos(q_1)+1)\cos(k_1) +  (\cos(q_2)+1)\cos(k_2) \\
  & +(\cos(q_1+q_2)+1)\cos(k_1+k_2) \\
  & +(\cos(\theta_{\vec{q}}) +\cos(\theta_{\vec{q}}-q_1)+\cos(\theta_{\vec{q}}-q_1-q_2))\\
  & -2(\cos(q_1)+\cos(q_2)+\cos(q_1+q_2))\\
  b(\vec{k})=& (\cos(q_1)-1)\cos(k_1) +  (\cos(q_2)-1)\cos(k_2) +\\
  & (\cos(q_1+q_2)-1)\cos(k_1+k_2) \\
  c(\vec{k})= & -\frac{1}{2}  ( (\cos(\theta_{\vec{q}})+1) +(\cos(\theta_{\vec{q}}-q_1)+1)e^{ik_1}\\
  & +(\cos(\theta_{\vec{q}}-q_1-q_2)+1)e^{i(k_1+k_2)} )\\
  d(\vec{k})=&-\frac{1}{2} ( (\cos(\theta_{\vec{q}})-1) +(\cos(\theta_{\vec{q}}-q_1)-1)e^{ik_1}\\
  & +(\cos(\theta_{\vec{q}}-q_1-q_2)-1)e^{i(k_1+k_2)} )\\
\end{align*}
The spin-wave dispersions $E_{\pm}^{\rm{SW}}(\vec{k})$ of Eq.~\eqref{bogol} can be obtained by solving the 
auxiliary eigenproblem 
\begin{equation}
  (\mathbf{A}(\vec{k})+\mathbf{B}(\vec{k}))(\mathbf{A}(\vec{k})-\mathbf{B}(\vec{k})) \Psi=\Big(E^{\rm{SW}}(\vec{k})\Big)^2 \Psi
\end{equation}
Solving this, we find that the  dispersions of the spin-wave modes are given by
\begin{align}
  E^{\rm{SW}}_{\pm}(\vec{k})=J^{\rm eff} \sqrt{\lambda_1(\vec{k}) \pm \lambda_2(\vec{k})},\\
  \lambda_1(\vec{k})=a(\vec{k})^2-b(\vec{k})^2+\lvert c(\vec{k})\rvert^2 -\lvert d(\vec{k})\rvert^2, \\
  \lambda_2(\vec{k})=\sqrt{4\lvert a(\vec{k})c(\vec{k})-b(\vec{k})d(\vec{k})\rvert^2 +(c(\vec{k})d(\vec{k})^{*}-c(\vec{k})^{*}d(\vec{k}))}
\end{align}
\section{Entropic selection of ground states at low temperatures}
\label{app_class}
As mentioned in Sec.~\ref{class_sel}, the matrix $K(\vec{q})$ , where $\vec{q}$ is the wave vector of the spiral ground state about which fluctuations are studied, is block diagonal in Fourier space.
The $2 \times 2$ blocks, labeled by the Fourier component $\vec{k}$ of the fluctuation, may be written as
\begin{align}
  \mathbf{K}(\vec{q},\vec{k})=J^{\rm eff}|S|^2
  \left(
  \begin{array}{cc}
    e(\vec{k})  & f^{*}(\vec{k}) \\
    f(\vec{k}) &  e(\vec{k}) \\
  \end{array}
  \right), 
\end{align}
where the explicit $\vec{q}$ dependence of $e$ and $f$ have been suppressed. The functions $e(\vec{k})$ and $f(\vec{k})$ are given by
\begin{align*}
  e(\vec{k})= & (\cos(q_1)+1)\cos(k_1) +  (\cos(q_2)+1)\cos(k_2) \\
  & +(\cos(q_1+q_2)+1)\cos(k_1+k_2) \\
  &+\frac{1}{2}(\cos(\theta_{\vec{q}}) +\cos(\theta_{\vec{q}}-q_1)+\cos(\theta_{\vec{q}}-q_1-q_2))\\
  f(\vec{k})=&  -\frac{1}{2}  \Big( \cos(\theta_{\vec{q}}) +\cos(\theta_{\vec{q}}-q_1)e^{ik_1}\\
  & +\cos(\theta_{\vec{q}}-q_1-q_2)e^{i(k_1+k_2)} \Big)\; .\\
\end{align*}
The eigenvalues of the $2\times2$ matrix $K(\vec{k},\vec{q})$ are given by $J^{\rm eff} |S|^2 \Big( e(\vec{k})\pm \lvert f(\vec{k})\rvert \Big)$.

\section{Details of Monte-Carlo Updates}
\label{MCapp}
Following Young {\em et. al.},~\cite{Lee_Young} we have used three different updates in our Monte Carlo:

a)\emph{Over-relaxation moves} : These are energy conserving moves where a spin  $\vec{S_i}$ is 
randomly selected, and reflected about local exchange field $\vec{H_i}$ induced by the coupling to other spins, with
\begin{equation}
  \vec{H_i}=\sum_j M_{ij}\vec{S_j}.
  \label{locF}
\end{equation}
This reflection is implemented by
\begin{equation}
  \vec{S_i} \rightarrow \vec{S_i}-\frac{2\vec{S_i} . \vec{H_i}}{\lvert \vec{H_i} \rvert}
\end{equation}
Over-relaxation moves help the simulations to equilibrate faster.

b)\emph{Heat-bath moves} : Over-relaxation moves described above are micro-canonical and therefore,
not ergodic. So, we supplement them with heat-bath moves. We randomly select a spin $\vec{S_i}$, and choose a new azimuthal angle $\theta$ and polar angle $\phi$ to specify its orientation relative to the local magnetic field $\vec{H_i}$ defined in Eq.~\eqref{locF}. The new angle $\theta$ is chosen with the \emph{heat-bath} probability $P(\cos(\theta))$, given by
\begin{equation}
  P(\cos(\theta))=\frac{\beta \vert\vec{H_i}\vert }{\sin(\beta \vert\vec{H_i} \vert)}
  \exp(-\beta \vert \vec{H_i}\vert \cos(\theta))
  \label{heatbath}
\end{equation}
As is well known, $cos(\theta)$ can be drawn from the above distribution by drawing a random number $r$ from a uniform distribution and equating it to the corresponding cumulative distribution. This prescription yields a random value for $\cos(\theta)$ in terms of the random number $r$:
\begin{equation}
  \cos(\theta) = -\frac{1}{\beta \vert \vec{H}\vert}\log\Big(1+r\exp(-2 \beta \vert \vec{H} \vert)\Big).
\end{equation}

If the azimuthal and polar angles made by the local field $\vec{H}$ with the co-ordinate axes are
$\theta'$ and$\phi'$ respectively, the spin with orientation $(\theta,\phi)$ with
respect to the effective magnetic field $\vec{H}$ can be written in our global coordinate system as:
\begin{align}
  \label{boring}
  S_{x} =&  \cos(\theta)  \sin(\theta')  \cos(\phi') \\
  &+ \sin(\theta)  \cos(\phi)  \sin(\phi')\\
  &+ \sin(\theta)  \sin(\phi)  \cos(\theta') \cos(\phi') \\
  S_{y} =& \cos(\theta)  \sin(\theta')  \sin(\phi') - \sin(\theta)  \cos(\phi)  \cos(\phi') \\
  & + \sin(\theta)  \sin(\phi)  \cos(\theta') \sin(\phi') \\
  S_{z} =& \cos(\theta') \cos(\theta) - \sin(\theta') \sin(\theta)  \sin(\phi) \\
\end{align}

c)\emph{Parallel Tempering} : Finally, we use parallel tempering or replica 
exchange to improve equilibration and eliminate loss of ergodicity at very low temperatures. We simultaneously run independent Monte Carlo simulations at a series of temperatures such that the highest few temperatures are high enough to not suffer from any loss of ergodicity. In a replica exchange move, one takes equilibriated configurations from independent simulations at $T_1$ and $T_2$ and exchanges the system configurations in their entirety between the two simulations, using an acceptance probability that obeys detailed balance

\begin{align}
  \label{ptemp}
  P(T_1 \leftrightarrow T_2)&=W_f/W_i, \text{ for } W_f<W_i\\ \nonumber
  &=1, \text{ otherwise}.  \\
 \end{align}
 The ratio of the weights $W_f/W_i$ is given in terms of the energies of the configurations $E_1$ and $E_2$ as

 \begin{equation}
  W_f/W_i= e^{-\frac{1}{T_1}(E_2-E_1)}e^{-\frac{1}{T_2}(E_1-E_2)}.
\end{equation}
Clearly, this is equivalent in practice to simply exchanging the temperatures of the two
independent simulations before restarting both of them, and this is what is
done in practice.


\begin{thebibliography}{999}
    \bibitem{Balents} L.~Balents, ``Spin liquids in frustrated magnets'', Nature {\bf 464}, 199-208 (2008).
    \bibitem{SavaryBalents} L.~Savary and L.~Balents, ``Quantum spin liquids'', arxiv 1601.03742 (2016).


    \bibitem{Sen_Damle} A.~Sen, K.~Damle and R.~Moessner, ``Fractional spin textures in the frustrated magnet ${\mathrm{SrCr}}_{9p}{\mathrm{Ga}}_{12-9p}{\mathrm{O}}_{19}$", Phys. Rev. Lett. {\bf 106}, 127203 (2011)

    \bibitem{Sen_Damle_PRB} A.~Sen, K.~Damle and R.~Moessner, ``Vacancy-induced spin textures and their interactions in a classical spin liquid", Phys. Rev. B {\bf 86}, 205134 (2012)

    \bibitem{Schiffer_Daruka} P.~Schiffer, I.~Daruka, ``Two-population model for anomalous low-temperature magnetism in geometrically frustrated magnets", Phys. Rev. B {\bf 56}, 13712 (1997)

    \bibitem{Limot} L.~Limot et al. ``Susceptibility and dilution effects of the kagom\'e bilayer geometrically frustrated network: A Ga NMR study of ${\mathrm{SrCr}}_{9p}{\mathrm{Ga}}_{12-9p}{\mathrm{O}}_{19}$", Phys. Rev. B. {\bf 65}, 144447 (2002)

    \bibitem{Lee} P.~A.~Lee, ``An end to the drought of quantum spin liquids", Science {\bf 321}, 1306–1307 (2008).

    \bibitem{Ramirez} A.~P.~Ramirez, ``Quantum spin liquids: A flood or a trickle?", Nature Phys. {\bf 4}, 442–443 (2008).
    \bibitem{Balz} C.~Balz et. al, ``Physical realization of a quantum spin liquid based on a complex frustration mechanism", Nat. Phys. {\bf 12}, 942-949 (2016).

    \bibitem{Balz2} C.~Balz et. al, ``The magnetic hamiltonian and phase diagram of the quantum spin liquid $\mathrm{Ca}_{10}\mathrm{Cr}_{7}\mathrm{O}_{28}$", Phys. Rev. B {\bf 95}, 174414 (2017).


    \bibitem{Moessner_Chalker_prl} R.~Moessner and J.~Chalker, ``Properties of a classical spin liquid: The Heisenberg pyrochlore antiferromagnet", Phys. Rev. Lett. {\bf 80}, 2929 (1998).

    \bibitem{Moessner_Chalker_prb} R.~Moessner and J.~Chalker, ``Low-temperature properties of classical geometrically frustrated antiferromagnets", Phys. Rev. B. {\bf 58}, 12049 (1998).

    \bibitem{Vesta} K.~Momma and F.~Izumi, ``Vesta 3 for three-dimensional visualization of crystal, volumetric and morphology data", J. App. Cryst. {\bf 44}, 1272-1276 (2011).

    \bibitem{Arun} A.~Mulder, R.~Ganesh, L.~Capriotti and A.~Paramekanti, ``Spiral order by disorder and lattice nematic order in a frustrated Heisenberg antiferromagnet on the honeycomb lattice",
      Phys. Rev. B {\bf 81}, 214419 (2010).

      \bibitem{Okumura} S.~Okumura, H.~Kawamura, T.~Okubo, and Y.~Motome, ``Novel spin-liquid states in the frustrated Heisenberg antiferromagnet on the honeycomb lattice'', J. Phys. Soc. Jpn. {\bf 79}, 114705 (2010).

    \bibitem{Rastelli} E.~Rastelli, A.~Tarsi and L.~Reatto , ``Non-simple magnetic order for simple Hamiltonians", Physica B, {\bf 97}, 1-24 (1979).

    \bibitem{Fouet} J.~B.~Fouet, P.~Sindzingre and C.~Lhuillier , ``An investigation of the quantum $J1-J2-J3$ model on the honeycomb lattice", Eur. Phys. J. B, {\bf 20}, 241-254 (2001).

    \bibitem{Mattsson}  A.~Mattsson, P.~Fr\"ojdh, and T.~Einarsson, "Frustrated honeycomb Heisenberg antiferromagnet: A Schwinger-boson approach", Phys.Rev. B. {\bf 49}, 3997 (1994).
    \bibitem{Arun2} R.~Ganesh, D.~N.~Sheng, Y.~J.~Kim and A.~Paramekanti,
    ``Quantum paramagnetic ground states on the honeycomb lattice and field-induced N\'eel order", Phys. Rev. B  {\bf 83}, 144414 (2011).


    \bibitem{PFFRG} M.~L.~Baez and J.~Reuther, ``Numerical treatment of spin systems with unrestricted spin length S: A functional renormalization group study", arXiv:1612.05074v1 (2016).


    \bibitem {Ma} S.~K.~Ma, ``Modern Theory of Critical Phenomena", Advanced Book
    Classics, Avalon Publishing (2000). 

    \bibitem{Garanin_Canals2} D.~A.~Garanin and B.~Canals, ``Spin-liquid phase in the pyrochlore anti-ferromagnet'', Can. J. Phy. {\bf 79} 1323-1331 (2001).

    \bibitem{Garanin_Canals1} D.~A.~Garanin and B.~Canals, ``Classical spin liquid: Exact solution for the infinite-component antiferromagnetic model on the kagome lattice'', Phy. Rev. B. {\bf 59} 443 (1999).

    \bibitem{Wang} F.~Wang, A.~Vishwanath and Y.~B.~Kim, ``Quantum and Classical spins on the spatially distorted Kagome lattice: applications to volborthite ${\mathrm{Cu}}_{3}{\mathrm{V}}_{2}{\mathrm{O}}_{7}{(\mathrm{O}\mathrm{H})}_{2}\dot{\mathrm{H}}_{2}\mathrm{O}$", Phys. Rev. B. {\bf 76}, 094421 (2007).

    \bibitem{Isakov} S.~V.~Isakov, K.~Gregor, R.~Moessner and S.~Sondhi, ``Dipolar Spin Correlations in Classical Pyrochlore Magnets", Phys. Rev. Lett. {\bf 93} 167204 (2004)

    \bibitem{Luttinger} J.~M.~Luttinger and L.~Tisza, ``Theory of dipole interaction in crystals'', Phys. Rev. {\bf70}, 954 (1946).

    \bibitem{Colpa} J.~H.~P.~Colpa, ``Diagonalization of the quadratic boson Hamiltonian with zero modes", Physica A,{\bf 2} 134, 417-422 (1986)

    \bibitem{Balents_diamond} D.~Bergman, J.~Alicea, E.~Gull, S.~Trebst and L.~Balents, ``Order-by-disorder and spiral spin-liquid in frustrated diamond-lattice antiferromagnets", Nat. Phys. {\bf 3}, 487-491 (2007).

    \bibitem{Wolf} U.~Wolff, ``Collective Monte Carlo updating for spin systems", Phys. Rev. Lett. {\bf 62}, 361 (1989)

    \bibitem{Hasenbusch} M.~Hasenbusch, ``Improved estimators for cluster updating of $O(n)$ spin models", Nuc. Phys. B. {\bf 333}, 581-592 (1990)

    \bibitem{Lee_Young} L.~W.~Lee and A.~P.~Young, ``Large-scale Monte Carlo simulations of the isotropic three-dimensional Heisenberg spin glass", Phys. Rev. B {\bf 76}, 024405 (2007).

    \bibitem{Pixley_Young} J.~H.~Pixley and A.~P.~Young, ``Large-scale Monte Carlo simulations of the three-dimensional XY spin glass", Phys. Rev. B {\bf 78}, 014419 (2008).

    \bibitem{Keren} A.~Keren, ``Dynamical simulation of spins on Kagom\'e and square lattices", Phys. Rev. Lett. {\bf 72} 3254 (1994).

    \bibitem{NicShannon} H.~Yan, R.~Pohle, and N.~Shannon, unpublished (N. Shannon, private communication).


\end{thebibliography}
\section{Form factors for $S=3/2$ moments}
\label{formfacs}
The form factors $f_1(\vec{k})$ and $f_2(\vec{k})$ for the effective $S=3/2$ degrees of freedom are given by
\begin{align}
  f_1(\vec{k})=&(1+e^{i k_1/2}+e^{-i k_2/2})e^{i(-k_1/6)+i(k_2/6)},\\
  \nonumber f_2(\vec{k})=&(1+e^{i k_1/2}+e^{i (k_1/2)+i(k_2/2)})e^{i(-k_1/3)+i(-k_2/6)}\\
  &\times e^{i(2 k_1/3)+i(k_2/3)} \times e^{i k_3 d_z},
\end{align}

where $d_z$ is the ratio of of the inter-layer separation  to the distance between two unit cells of the triangular Bravais lattice of the effective model. This number is never actually needed for our purposes because we calculate momentum dependent quantities like structure factors with the out of plane momentum $k_3$ set to zero.

\end{document}